\documentclass[12pt]{article}

\PassOptionsToPackage{nosort}{cite}

\usepackage{amssymb}
\usepackage{amsmath}
\usepackage{trimclip}
\usepackage{mathtools}
\usepackage{dsfont}
\usepackage{mathrsfs}
\usepackage{enumitem}
\usepackage[T1]{fontenc}
\usepackage{chet}
\usepackage[justification=centering]{caption}
\usepackage{float}
\usepackage{tikz}
\usepackage{hyperref}
\usepackage{verbatim}

\usetikzlibrary{shapes.misc,arrows,decorations.markings,patterns,calc,shapes.geometric}

\definecolor{darkblue}{rgb}{0.1,0.1,0.7}
\hypersetup{colorlinks,
           linkcolor={darkblue},
           citecolor={darkblue},
           urlcolor={darkblue},
           pdftitle={Transdimensional Defects}
           pdfdisplaydoctitle
}

\newcommand{\lsp}{\hspace{0.5pt}}

\renewcommand{\geq}{\geqslant}
\renewcommand{\leq}{\leqslant}
\newcommand{\del}{\partial}
\newcommand{\lf}{\left}
\newcommand{\rg}{\right}
\newcommand{\xp}{x_\perp}
\newcommand{\yp}{y_\perp}
\newcommand{\hphi}{{\hat{\phi}}}

\newcommand{\hD}{\hat{\Delta}}

\newcommand{\veps}{\varepsilon}

\def\Nm{{\mathcal{N}}}

\DeclarePairedDelimiter{\cor}{\langle}{\rangle}

\allowdisplaybreaks

\title{Transdimensional Defects}
\author{
\begin{tabular}{ccc}
\large Elia de Sabbata,${\!}^{a,c}$ &
\large Nadav Drukker,${\!}^c$ &
\large Andreas Stergiou$^{c}$
\\[1mm]
\footnotesize\hyperlink{mailto:eliadesabbata@gmail.com}{eliadesabbata@gmail.com} &
\footnotesize\hyperlink{mailto:nadav.drukker@gmail.com}{nadav.drukker@gmail.com} &
\footnotesize\hyperlink{mailto:andreas.stergiou@kcl.ac.uk}{andreas.stergiou@kcl.ac.uk}
\end{tabular}
}

\affiliation{%
$^a$Dipartimento di Scienze e Innovazione Tecnologica, Universit\`a del
Piemonte Orientale,\\
Via T. Michel 11, 15121 Alessandria, Italy,\\
and INFN - Sezione di Torino
Via P. Giuria 1, 10125 Torino, Italy\\[2mm]
$^b$Dipartimento di Fisica, Universit\`a di Torino and INFN - Sezione di Torino\\
Via P. Giuria 1, 10125 Torino, Italy\\[2mm]
$^c$Department of Mathematics, King's College London,\protect\\
Strand, London, WC2R 2LS, United Kingdom}

\abstract{This note introduces a novel paradigm for conformal defects with continuously adjustable dimensions. Just as the standard $\varepsilon$ expansion interpolates between integer spacetime dimensions, a new parameter, $\delta$, is used to interpolate between different integer-dimensional defects. This framework is explored in detail for defects of dimension $p=2+\delta$ in both free and interacting $O(N)$ bulk conformal field theories (CFTs) in $d=4-\varepsilon$. Comprehensive calculations are performed to first and second order in $\varepsilon$ and to high or all orders in $\delta$. Additionally, in the large-$N$ limit, the interpolation between defects of dimensions $p=1$ and $p=2$ is analysed for spacetime dimensions $4\leq d\leq 6$. The new parameter $\delta$ provides a natural enrichment of the space of defect CFTs and allows to find new integer dimension or co-dimension defects.}

\date{November 2024}

\begin{document}

\maketitle

\section{Introduction}\label{sec:intro}
The influence of the Wilson--Fisher dimensional continuation on the study of fixed points and the renormalisation group (RG) \cite{Wilson:1971dc} cannot be overstated. Their versatile approach, which is recognised by the name ``$\veps$ expansion,'' allows for good approximate control of many strongly coupled conformal field theories (CFTs), including the Ising and $O(N)$ models in $d=3$~\cite{Kompaniets:2017yct}, and also allows to formulate non-trivial RG flows and CFTs in dimension ``$d=3.99$''~\cite{Pelissetto:2000ek, Osborn:2017ucf}.

In recent years, there is an explosion of interest in conformal defects in CFTs (or dCFT), i.e.~systems with conformal symmetry along a $p$-dimensional submanifold of space. This includes many examples of conformal defects in the $\veps$ expansion, mostly the critical $O(N)$ model, but also systems with other symmetries. For a partial list of references see~\cite{McAvity:1995zd, Billo:2016cpy, Cuomo:2021kfm, Gimenez-Grau:2022czc, Giombi:2022vnz, Bianchi:2022sbz, Gimenez-Grau:2022ebb, Pannell:2023pwz, Trepanier:2023tvb, Raviv-Moshe:2023yvq, Giombi:2023dqs, Bianchi:2023gkk, Harribey:2023xyv, Dey:2024ilw, Harribey:2024gjn, Pannell:2024hbu, Bashmakov:2024suh}. To the extent of our knowledge, all the many examples studied thus far have had either integer $p=1,2$ or integer $q=d-p=1,2$. The former are line or surface defects and remain lines or become interfaces as $\veps\to1$. The latter are boundaries/interfaces or vortex-like configurations and retain those properties for all $\veps$. The purpose of this paper is to relax this assumption and allow for arbitrary $p$ with a focus on ``the $p=2.01$ defect in $d=3.99$ dimensions.''

This idea immediately implies some variation on the already known examples of conformal defects. The additional parameter $\delta$ in $p=2+\delta$ allows for new perturbative expansions. It also allows for new interpolations, for example $p:2\to3$ as $d:4\to3$, giving new bulk three-dimensional CFTs, or $p:2\to3-\veps$ with fixed $d=4-\veps$, giving possibly new interfaces. These are the focus of the next section, while in Section~\ref{sec:largeN} we take $p:1\to2$ as $d:4\to6$.

This summarises the main idea of the paper. The next sections illustrate it in the examples mentioned. We perform all-order (or high-order) calculations in $\delta$ at fixed order in $\veps$ or at large $N$ that enable real interpolation in the defect dimension. Each section is self-contained and the results of those calculations and multitude of possible generalisations are presented in the discussion section.

\section{Defects of dimension
\texorpdfstring{$\boldsymbol{p=2+\delta}$ in $\boldsymbol{d=4-\veps}$}{p=2+delta in d=4-epsilon}}
\label{sec:2+delta}

\subsection{Free \texorpdfstring{$O(N)$}{O(N)} field theory}
\label{sec:free}

We begin with a simple soluble example, the free scalar $O(N)$ theory in $d$ dimensions. Its action is
\begin{equation}\label{freedefectaction}
 S=\int d^{\lsp d} x \, \tfrac12\lsp \partial^\mu \phi_a\lsp\partial_\mu\phi_a\,.
\end{equation}
We deform the action by the defect interaction terms
\begin{equation}
\label{STaction}
 S_{D}= \tfrac{1}{2}(h_0)_{ab}\int d^{\lsp p} \tau \, ( \delta_{ab} S + T_{ab} )\,,
\end{equation}
where $S$ and $T_{ab}$ are the two primary operators that are quadratic in the fields,
\begin{equation}
\label{ST}
S \equiv \frac{1}{N}\phi_a \phi_a\,, \qquad
T_{ab} \equiv \phi_a \phi_b- \frac{\delta_{ab}}{N}\phi_c \phi_c\,,
\end{equation}
and $h_0$ has mass dimension $p+2-d$. For simplicity we focus on the $O(N)$-invariant defect.

For $d=4-\veps<4$ and $p=2$, this deformation is weakly relevant and triggers a defect RG flow. This flow can be studied exactly in $\veps$, and for $\veps=1$ it reaches an infrared (IR) fixed point that corresponds to the three-dimensional free scalar theory with an interface with Dirichlet boundary conditions~\cite{Trepanier:2023tvb, Raviv-Moshe:2023yvq, Giombi:2023dqs}. If we instead take $d=p=3$, the action just defines a free massive scalar theory with mass $m^2=h_0$. Here we explore a third possibility by setting $d=4-\veps$ and $p=2+\delta$. The deformation is again weakly relevant and the defect flows to an IR fixed point. This defect RG flow can be studied in perturbation theory, where it is possible to resum diagrams at all orders. The analysis is essentially the same as in~\cite{Trepanier:2023tvb, Giombi:2023dqs}, except that defect integrals have now to be performed in $p=2+\delta$ dimensions. A similar approach has been explored in the context of coupling theories across different dimensions~\cite{Teber:2012de}.

More concretely, the one-point function is given by the sum of the series of diagrams
\begin{equation}\label{diagram1}
\begin{tikzpicture}[scale=0.5,baseline=(vert_cent.base)]
  \draw[very thick,black] (-2,0)--(2,0);
  \draw[thick, dashed] (0,3) to[out=-45,in=45](0,0);
	 \draw[thick, dashed] (0,3) to[out=-135,in=135](0,0);
  \fill[black] (0,0) circle (4pt);
	 \node[above] at (0,3) {$\phi^2$};
  \node[inner sep=0pt,outer sep=0pt] (vert_cent) at (0,1.75) {$\phantom{\cdot}$};
\end{tikzpicture}
\ +\
\begin{tikzpicture}[scale=0.5,baseline=(vert_cent.base)]
	 \draw[very thick,black] (-3,0)--(3,0);
	 \draw[thick, dashed] (0,3)--(1.5,0);
	 \draw[thick, dashed] (0,3)--(-1.5,0);
  \draw[thick, dashed] (-1.5,0)to[out=60,in=120](1.5,0);
  \fill[black] (-1.5,0) circle (4pt);
  \fill[black] (1.5,0) circle (4pt);
	 \node[above] at (0,3) {$\phi^2$};
  \node[inner sep=0pt,outer sep=0pt] (vert_cent) at (0,1.75) {$\phantom{\cdot}$};
\end{tikzpicture}
\ +\
\begin{tikzpicture}[scale=0.5,baseline=(vert_cent.base)]
	 \draw[very thick,black] (-3,0)--(3,0);
	 \draw[thick, dashed] (0,3)--(1.5,0);
	 \draw[thick, dashed] (0,3)--(-1.5,0);
  \draw[thick, dashed] (0,0)to[out=100,in=80](-1.5,0);
  \draw[thick, dashed] (0,0)to[out=80,in=100](1.5,0);
  \fill[black] (-1.5,0) circle (4pt);
  \fill[black] (0,0) circle (4pt);
  \fill[black] (1.5,0) circle (4pt);
  \node[above] at (0,3) {$\phi^2$};
  \node[inner sep=0pt,outer sep=0pt] (vert_cent) at (0,1.75) {$\phantom{\cdot}$};
\end{tikzpicture}
\ +\ \cdots
\end{equation}
Here the solid line represents the $p$-dimensional defect and dashed lines are free propagators
\begin{equation}\label{eq:freeprop}
G(x)=\frac{\Nm_\phi^{\lsp 2}}{|x|^{d-2}}\,, \qquad
\Nm_\phi^{\lsp 2}=\frac{\Gamma(d/2-1)}{4 \pi^{d/2}}\,.
\end{equation}
to renormalise the defect we set $h_0=\mu^{\veps+\delta}Z_h  h$, where $Z_h$ is a renormalisation factor and $h$ is the renormalised coupling. We then impose finiteness of the bulk one-point function $\cor{\phi^2 (\xp,x_\parallel)}$ in the limit $\veps,\delta \rightarrow 0$, where $\xp$ is the distance from the defect. In the MS scheme,%
\footnote{
In this case there are only poles in the variable $\veps+\delta$, hence the MS scheme is unambiguous. In the interacting case, where the poles are different linear combinations of $\veps$ and $\delta$, we need to specify how the minimal subtraction should be performed.} finiteness of the sum in \eqref{diagram1} is guaranteed at all orders provided that
\begin{equation}\label{Zrenfreebulk}
    Z_h=\frac{1}{1-h/(\veps+\delta)}\,,
\end{equation}
where we rescaled $h\to2\pi h$. The exact beta function is obtained by imposing $\mu \frac{d}{d\mu} h_0=0$; it reads
\begin{equation}\label{eq:betahfree}
    \beta_h=-(\veps+\delta)h+h^2\,,
\end{equation}
and admits a non-trivial IR fixed point for $h_*=\veps+\delta$. The special case of $\veps+\delta=0$ is mentioned in the discussion section.

Since the action \eqref{freedefectaction} is Gaussian, generic defect correlators can be obtained by Wick contractions, once we know the dimension of the defect operator $\hphi$, $\smash{\hD_{\hphi}}$, and that of the defect operator $\partial_\perp\hphi$, $\smash{\hD_{\partial_\perp\hphi}}$. (Throughout this work we follow the convention of dressing defect observables and operators with a hat.) To determine the first, we use that at the fixed point
\begin{equation}\label{freedefectdimensions}
 \hD_{\hat S}=p+\frac{\del \beta_h}{\del h}\Big|_{h_*}=2+\veps+2\lsp\delta
 \qquad\Rightarrow\qquad
 \hD_\hphi=1+\frac{\veps}{2}+\delta\,.
\end{equation}
The operator $\del_\perp \hphi$ does not get renormalised in this model, see Footnote~\ref{foot:deriv}, and its dimension is
\begin{equation}
    \hat{\Delta}_{\del_\perp \hphi}=2-\frac{\veps}{2}\,.
\end{equation}

The sum in \eqref{diagram1} also determines the value of the one-point function of $S$ at the bulk fixed point in the presence of the defect, see (\ref{onepointallorderapp}, \ref{bkcoeff}). To compute it, we expand the result in powers of $\veps$ and $\delta$ which allows us to guess the expression
\begin{equation}\label{freeonepoint}
\begin{split}
\langle S (\xp,x_\parallel) \rangle &= \frac{\mathcal{N}_{S}\, a_{S}}{|\xp|^{2\Delta_\phi}}\,, \qquad
\mathcal{N}_{S}=2 N\lsp \mathcal{N}_\phi^{\,2}\,,
\qquad
a_{S}= -\frac{\sqrt{\pi}\,\Gamma\! \left( 1+\frac{\veps}{2}+\delta \right)}{2^{2+\delta}\lsp\Gamma\!\left( \frac{3+\delta}{2}\right)\Gamma\!\left(\frac{\veps +\delta}{2} \right)}\,.
\end{split}
\end{equation}

With these expressions in hand, we analyse the system for different values of $\delta$. For $\delta=1-\veps$ the defect is an interface and \eqref{freedefectdimensions}, \eqref{freeonepoint} reproduce, for any $\veps$, the values for the free scalar theory with Dirichlet boundary conditions \cite{McAvity:1995zd}. The dimensions of both $\partial_\perp\hphi$ and $\hphi$ are $2-\veps/2$ and further evidence that they are linearly dependent comes from the boundary operator product expansion (OPE) coefficients of these two operators \cite{Billo:2016cpy}. At order $\veps^0$,
\begin{equation}\label{bOPE coefficients}
\langle \phi(x) \hphi(\tau)\rangle = \frac{\Nm_\phi \lsp \Nm_\hphi \lsp b_{\phi , \hphi}}{|x_\perp|^{\Delta_\phi-\hat{\Delta}_\hphi}|x_\parallel-\tau|^{2\hat{\Delta}_\hphi}}\,,
\qquad
b^{\lsp 2}_{\phi, \hphi}= \frac{2^\delta \lsp \Gamma\!\left( 1-\frac{\delta}{2}\right) \Gamma\!\left( \frac{1+\delta}{2}\right)}{\sqrt{\pi}}+\text{O}(\veps)\,,
\end{equation}
where $\langle \hphi(0) \hphi(\tau)\rangle= \Nm_\hphi^{\lsp 2}/|\tau|^{2\hat{\Delta}_\hphi}$, and similarly
\begin{equation}
  b^{\lsp 2}_{\phi,\del_\perp \hphi}=\frac{2}{2-\delta}+\text{O}(\veps)\,.
\end{equation}
This means that for $\veps=0$ (and $\delta=1$), the boundary OPE coefficient for the unit normalised operator $\hat{\Phi}_+=(\hphi+\del_\perp \hphi)/\sqrt{2}$ becomes $ b^{\lsp 2}_{\phi, \hat{\Phi}_+}=( b_{\phi, \hphi}+b_{\phi, \del_\perp\hphi})^2/2=4$, which is the correct value for the boundary operator in the Dirichlet interface theory in $d=4$. Similarly, we can see that the operator $\hat{\Phi}_-=(\hphi-\del_\perp \hphi)/\sqrt{2}$ decouples. A similar statement should hold for $\veps>0$.

As another example, $\delta \rightarrow -1$ should lead to a line defect. To keep the quadratic deformation relevant, we must start with $\veps+\delta>0$. For the case of $\veps=1$, we get $\hD_\hphi=1/2$ from \eqref{freedefectdimensions}, which is the dimension of the bulk field, so this is the trivial defect with $h_*=0$. A similar result was obtained for $N=1$ in~\cite{Lauria:2020emq}.

Next, for $\veps,\delta \rightarrow 1$ we get a three-dimensional theory with $\hD_{\hphi}=5/2$, so a non-local theory of generalised free fields (GFF). More generally, leaving $0<\veps<1$ as a free parameter and looking only at defect correlators, one finds a continuous family of GFF theories with $2<\hD_{\hphi}<5/2$. This is similar in spirit to the defect description of the long-range Ising model~\cite{Paulos:2015jfa}. Another generalisation, starting with any theory near the critical dimension $d_c$, i.e. $d=d_c-\veps$ and $p=d_c-2+\delta$ for some positive integer $d_c\geq2$, gives $\hat{\Delta}_{\hat{\phi}}=\frac12(d_c+1)$ when $\veps,\delta\to 1$.

Finally, it is instructive to see what happens in the space-filling limit $p \to d$, i.e.\ $\delta\to2-\veps$. Since in this limit there are no orthogonal directions ($|x_\perp|\to0$), the most one can do is push bulk operators to the defect by suitable rescalings. This procedure selects the lightest defect operator in the defect OPE. For example, in this limit the unit normalised operator $b^{-1}_{S, (\del_\perp\hphi)^2}|\xp|^{\Delta_{S}-\hat{\Delta}_{(\del_\perp\hphi)^2}} S(x)$ reduces to the defect operator $(\del_\perp\hphi)^2(x)$. Note that the dimensions of $\hphi$ and $\del_\perp \hphi$ cross at the interface value $\delta=1-\veps$ and for $\delta>1-\veps$ the latter becomes smaller. Crucially, for $\delta=2-\veps$ the bulk one-point function coefficient $a_{S}$ remains finite, whereas the defect OPE coefficient
\begin{equation}
 b_{S, (\del_\perp\hphi)^2}= \frac{8N\lsp\pi^\veps\lsp \Gamma\!\left(2-\frac{\veps}{2} \right)^2}{(2-\veps-\delta)^2}\,,
\end{equation}
is divergent. Therefore, $\lim_{p \to d}a_{S}/b_{S , (\del_\perp\hphi)^2}=0$, so that there are no non-zero one-point functions.

\subsection{The interacting \texorpdfstring{$O(N)$}{O(N)} vector model}
\label{sec:interacting}

After studying the free theory, we are ready to tackle the interacting $O(N)$ model with action
\begin{equation}\label{ondefectaction}
 S=\int d^{\lsp d} x \, \big[ \tfrac12\lsp \partial^\mu \phi_a\lsp\partial_\mu\phi_a +\tfrac{1}{8}\lambda_0\lf( \phi_a \phi_a\rg)^2 \big] .
\end{equation}
For $d=4-\veps<4$ the coupling triggers an RG flow. The one-loop beta function for $\lambda$ (after rescaling $\lambda\to 16\pi^2 \lambda$) is
\begin{equation}\label{bulkbeta}
 \beta_\lambda=-\veps \lambda + (N+8)\lambda^2+\text{O}(\lambda^3)\,.
\end{equation}
This flow ends at the IR Wilson--Fisher fixed point at
\begin{equation}\label{wilsonfisherfixedpoint}
 \lambda_*=\frac{\veps}{N+8} + \text{O}(\veps^2)\,.
\end{equation}

The conformal dimensions of the $S$ and $T_{ab}$ operators in \eqref{ST} are (see~\cite{Kleinert:2001ax, Henriksson:2022rnm} for a variety of methods and results)
\begin{equation}
\Delta_S=2-\frac{6}{N+8}\veps+\text{O}(\veps^2)\,,
\qquad
\Delta_T=2-\frac{N+6}{N+8}\veps+\text{O}(\veps^2)\,.
\end{equation}
The defect action \eqref{STaction} then triggers a defect RG flow since the coupling $\lf(h_0\rg)_{ab}$ is relevant for any $N$ provided that $p\gtrsim 2$.

Focusing again on the symmetry preserving defect with $(h_0)_{ab}=h_0\delta_{ab}$, the diagrams contributing to the one-point function $\langle S\rangle$ at first order in $\lambda$ and at all orders in $h$ are
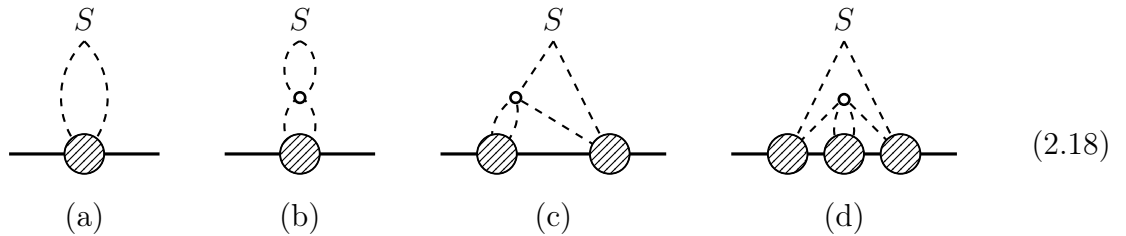
\begin{equation}\label{diagrams1sc}
\begin{tikzpicture}[scale=0.5, baseline= 0 cm]
	 \draw[very thick,black] (-2,0)--(2,0);
	 \draw[thick, dashed] (0,3) to[out=-45,in=45](0,0);
	 \draw[thick, dashed] (0,3) to[out=-135,in=135](0,0);
 \draw[fill= white] (0,0) circle (15pt);
 \draw[fill= white, draw=black, line width=0.25mm, pattern=north east lines] (0,0) circle (15pt);
 \node[above] at (0,3) {$S$};
 \node[below] at (0,-1) {(a)};
\end{tikzpicture}
\qquad
\begin{tikzpicture}[baseline=0cm,scale=0.5]
 \draw[very thick,black] (-2,0)--(2,0);
 \draw[thick, dashed] (0,3) to[out=-20,in=20](0,1.5);
 \draw[thick, dashed] (0,3) to[out=-160,in=160](0,1.5);
 \draw[thick, dashed] (0,1.5) to[out=-20,in=20](0,0);
 \draw[thick, dashed] (0,1.5) to[out=-160,in=160](0,0);
 \filldraw[color=black,fill=white,line width=0.35mm] (0,1.5) circle (4pt);
 \draw[fill= white] (0,0) circle (15pt);
 \draw[fill= white, draw=black, line width=0.25mm, pattern=north east lines] (0,0) circle (15pt);
 \node[above] at (0,3) {$S$};
 \node[below] at (0,-1) {(b)};
\end{tikzpicture}
 \qquad
\begin{tikzpicture}[scale=0.5, baseline= 0 cm]
	 \draw[very thick,black] (-3,0)--(3,0);
	 \draw[thick, dashed] (0,3)--(1.5,0);
	 \draw[thick, dashed] (0,3)--(-1,1.5);
 \draw[thick, dashed] (1.5,0)to[out=150,in=-30](-1,1.5);
 \draw[thick, dashed] (-1.5,0)to[out=120,in=210](-1,1.5);
 \draw[thick, dashed] (-1.5,0)to[out=45,in=285](-1,1.5);
 \filldraw[color=black,fill=white,line width=0.35mm] (-1,1.5) circle (4pt);
 \draw[fill= white] (-1.5,0) circle (15pt);
 \draw[fill= white, draw=black, line width=0.25mm, pattern=north east lines] (-1.5,0) circle (15pt);
 \draw[fill= white] (1.5,0) circle (15pt);
 \draw[fill= white, draw=black, line width=0.25mm, pattern=north east lines] (1.5,0) circle (15pt);
	 \node[above] at (0,3) {$S$};
 \node[below] at (0,-1) {(c)};
\end{tikzpicture}
 \qquad
 \begin{tikzpicture}[scale=0.5, baseline= 0 cm]
	 \draw[very thick,black] (-3,0)--(3,0);
	 \draw[thick, dashed] (0,3)--(1.5,0);
	 \draw[thick, dashed] (0,3)--(-1.5,0);
 \draw[thick, dashed] (-1.5,0)to(0,1.5);
 \draw[thick, dashed] (1.5,0)to(0,1.5);
 \draw[thick, dashed] (0,1.5)to[out=-60,in=45](0,0);
 \draw[thick, dashed] (0,1.5)to[out=240,in=135](0,0);
 \draw[fill= white] (1.5,0) circle (15pt);
 \draw[fill= white, draw=black, line width=0.25mm, pattern=north east lines] (1.5,0) circle (15pt);
 \draw[fill= white] (0,0) circle (15pt);
 \draw[fill= white, draw=black, line width=0.25mm, pattern=north east lines] (0,0) circle (15pt);
 \filldraw[color=black,fill=white,line width=0.35mm] (0,1.45) circle (4pt);
 \draw[fill= white] (-1.5,0) circle (15pt);
 \draw[fill= white, draw=black, line width=0.25mm, pattern=north east lines] (-1.5,0) circle (15pt);
	 \node[above] at (0,3) {$S$};
 \node[below] at (0,-1) {(d)};
	\end{tikzpicture}
\end{equation}
Here the bubbles represent contributions at all powers of $h$, or ``hops'' on the defect.  We first define them recursively as an effective defect-to-defect propagator
\begin{equation}\label{effproprecursion}
 \begin{tikzpicture}[baseline=(vert_cent.base),scale=0.5]
 \draw[very thick,black] (-3.5,0)--(3.5,0);
 \draw[thick, dashed] (-3,0)to[out=90,in=180](-1.5,1.5)to[out=0,in=90](0,0);
 \draw[thick, dashed] (3,0)to[out=90,in=0](1.5,1.5)to[out=180,in=90](0,0);
 \draw[fill= white] (0,0) circle (15pt);
 \draw[fill= white, draw=black, line width=0.25mm, pattern=north east lines] (0,0) circle (15pt);
 \node[inner sep=0pt,outer sep=0pt] (vert_cent) at (0,0) {$\phantom{\cdot}$};
\end{tikzpicture}
 \ = \
 \begin{tikzpicture}[baseline=(vert_cent.base),scale=0.5]
 \draw[very thick,black] (-2.5,0)--(2.5,0);
 \draw[thick, dashed, shorten <=2] (-1.5,0)to[out=90,in=180](0,1.5)to[out=0,in=90](1.5,0);
 \node[inner sep=0pt,outer sep=0pt] (vert_cent) at (0,0) {$\phantom{\cdot}$};
\end{tikzpicture}
\ + \
 \begin{tikzpicture}[baseline=(vert_cent.base),scale=0.5]
 \draw[very thick,black] (-6.5,0)--(3.5,0);
 \draw[thick, dashed, shorten <=1] (0,0)to[out=90,in=0](-1.5,1.5)to[out=180,in=90](-3,0);
 \draw[thick, dashed, shorten <=1] (0,0)to[out=90,in=180](1.5,1.5)to[out=0,in=90](3,0);
 \draw[fill= white] (0,0) circle (15pt);
 \draw[fill= white, draw=black, line width=0.25mm, pattern=north east lines] (0,0) circle (15pt);
 \fill[black] (-3,0) circle (4pt);
 \draw[thick, dashed, shorten <=1] (-6,0)to[out=90,in=180](-4.5,1.5)to[out=0,in=90](-3,0);
 \node[inner sep=0pt,outer sep=0pt] (vert_cent) at (0,0) {$\phantom{\cdot}$};
 \end{tikzpicture}
 \ \,.
\end{equation}
This is evaluated as an infinite sum in Appendix~\ref{appintegrals}, see \eqref{effpropapp}. From this we define the contraction with the bulk
\begin{equation}\label{bulkeffproprecursion}
 \begin{tikzpicture}[scale=0.5, baseline=(vert_cent.base)]
	 \draw[very thick,black] (-2,0)--(2,0);
	 \draw[thick, dashed] (1.5,2.5)to (0,0);
	 \draw[thick, dashed] (-1.5,2.5)to (0,0);
  \draw[fill= white] (0,0) circle (15pt);
  \draw[fill= white, draw=black, line width=0.25mm, pattern=north east lines] (0,0) circle (15pt);
  \node[inner sep=0pt,outer sep=0pt] (vert_cent) at (0,0) {$\phantom{\cdot}$};
\end{tikzpicture}
 \ \equiv \
\begin{tikzpicture}[scale=0.5,baseline=(vert_cent.base)]
	 \draw[very thick,black] (-2,0)--(2,0);
	 \draw[thick, dashed] (1.5,2.5)to (0,0);
	 \draw[thick, dashed] (-1.5,2.5)to (0,0);
  \fill[black] (0,0) circle (4pt);
  \node[inner sep=0pt,outer sep=0pt] (vert_cent) at (0,0) {$\phantom{\cdot}$};
\end{tikzpicture}
 \ + \
\begin{tikzpicture}[scale=0.5,baseline=(vert_cent.base)]
	 \draw[very thick,black] (-3,0)--(3,0);
	 \draw[thick, dashed] (2,0)--(2,2.5);
	 \draw[thick, dashed] (-2,0)--(-2,2.5);
  \fill[black] (-2,0) circle (4pt);
  \fill[black] (2,0) circle (4pt);
  \draw[thick, dashed] (-2,0)to[out=90,in=180](-1,1)to[out=0,in=90](0,0);
  \draw[thick, dashed] (2,0)to[out=90,in=0](1,1)to[out=180,in=90](0,0);
  \draw[fill= white] (0,0) circle (15pt);
  \draw[fill= white, draw=black, line width=0.25mm, pattern=north east lines] (0,0) circle (15pt);
  \node[inner sep=0pt,outer sep=0pt] (vert_cent) at (0,0) {$\phantom{\cdot}$};
\end{tikzpicture}
 \ \,.
\end{equation}
The expression for this is more complicated and we do not have it in closed form for arbitrary bulk endpoints. For the purpose of evaluating the divergences in the diagrams~\eqref{diagrams1sc}, we only need it in two cases: with two coincident endpoints and with one endpoint close to the defect and the other far. These are written in \eqref{onepointallorderapp} and (\ref{cd}-\ref{dkcoeff}).

At the end of the calculation of the divergences in~\eqref{diagrams1sc}, we write the counterterm as an extra factor multiplying $Z_h \big|_{\lambda=0}$ of the free theory in \eqref{Zrenfreebulk} and its first terms are
\begin{equation}\label{hcounterterms}
 Z_{h,\lambda}= Z_h \big|_{\lambda=0} \bigg( 1+ \lambda \, (N+2) \Big(\frac{1}{\varepsilon}+\frac{h}{\left( \varepsilon + \delta \right)\left( 2\varepsilon + \delta \right)}-\frac{2h}{2\varepsilon +\delta}+ \cdots\Big)\bigg) + \text{O}(\lambda^2)\,.
\end{equation}
As usual the relation between the bare and the renormalised coupling is $h_0 = \mu^{\veps+\delta} Z_{h,\lambda} \, h$. Then imposing that $h_0$ does not depend on the scale $\mu$, one gets the beta function
\begin{equation}\label{interactingbeta}
 \beta_h=-( \varepsilon+\delta)h +h^2+\lambda \, (N+2) \left(h-2h^2+2h^3+\cdots \right)+\text{O}(\lambda^2)\,,
\end{equation}
with terms up to order $\lambda\lsp h^{11}$ written in \eqref{fullbetah}. From this one can compute the fixed point coupling $h_*$ \eqref{shorth*} and the dimension of the defect operator $\hat{S}$ at the fixed point is
\begin{equation}\label{deltaphi2}
 \hD_{\hat{S}}= p+\frac{\del \beta_h}{\del h}\Big|_{\lambda_*,h_*}\!\!= 2 + 2 \delta +\varepsilon \left( \frac{6}{N+8}+ \frac{N+2}{N+8} \, f(\delta) \right)+ \text{O}(\varepsilon^2)\,,
\end{equation}
where the first ten terms in the expansion of $f(\delta)$ can be conveniently packaged as
\begin{equation}
\begin{aligned}
f(\delta)= \frac{\delta-1}{\delta+1}+\frac{1}{1+2\delta}\exp \Big( \zeta(3)\lsp\delta^3 -\frac{9}{4} \zeta(4)\lsp\delta^4 +\frac{9}{2} \zeta(5) \lsp\delta^5-\frac{135}{16} \zeta(6)\lsp\delta^6
+\frac{249}{16} \zeta(7) \lsp\delta^7
\\
-\frac{1827}{64} \zeta(8)\lsp\delta^8 +\frac{2515}{48} \zeta(9)\lsp\delta^9
-\frac{24687}{256} \zeta(10) \lsp\delta^{10} + \text{O}(\delta^{11})\Big).
\label{asymptoticseries}
\end{aligned}
\end{equation}
While this is an economical way to write the answer, we have not found a pattern in the rational coefficients multiplying the zeta values and the terms in the exponent do not decrease, so it does not provide a good indication of the analytic structure of $f(\delta)$. Instead, if we expand everything in a power series, the ratio of subsequent terms $f_k/f_{k+1}\xrightarrow{\sim}-2/3$. This indicates a pole and we can then repeat the analysis with the function $(\delta+2/3)f(\delta)$, whose series suggests a further singularity at $\delta=-1$. Finally, the series of the function $(\delta+1)(\delta+2/3)f(\delta)$ is now quickly convergent for $|\delta| \leq 1$ and can be used to estimate both the value and the error of $f(1)$.\footnote{A conservative way to estimate the error is to first observe that the ratios of consecutive coefficients of $g(\delta)$ are decreasing. Then the error can be computed as a geometric series starting with the last known coefficient and with rate given by the last known ratio.} In this way, we find
\begin{equation}
 f(1) = 0.4999(3)\,.
\end{equation}

\begin{figure}[ht]
\centering
\begin{tikzpicture}[scale=8, baseline=0 cm]
\draw[] (-0.3,0) -- (1,0) node[right] {$\delta$};
\draw[] (0,0) -- (0,0.5) node[above=5pt] {$f(\delta)$};
\foreach \x/\xtext in {-0.25/-0.25, 0/0, 0.25/0.25, 0.5/0.5, 0.75/0.75, 1/1}
\draw[shift={(\x,0)}] (0pt,.3pt) -- (0pt,-.3pt) node[below] {$\xtext$};
\foreach \y/\ytext in {0.25/0.25, 0.5/0.5}
\draw[shift={(0,\y)}] (0.3pt,0pt) -- (-0.3pt,0pt) node[left] {$\ytext$};

\draw[color=black, thick]
plot coordinates {(-0.31, 0.508237) (-0.309598, 0.506066) (-0.309196, 0.503904) (-0.308393, 0.499605)
(-0.306786, 0.491107) (-0.303572, 0.474498) (-0.297143, 0.442782)
(-0.296741, 0.440864) (-0.29634, 0.438953) (-0.295536, 0.435153)
(-0.293929, 0.427642) (-0.290715, 0.41296) (-0.284286, 0.384918)
(-0.283851, 0.383079) (-0.283415, 0.381248) (-0.282544, 0.377609)
(-0.280802, 0.370421) (-0.277317, 0.356399) (-0.270348, 0.329716)
(-0.269913, 0.328106) (-0.269477, 0.326503) (-0.268606, 0.323318)
(-0.266864, 0.317025) (-0.263379, 0.304749) (-0.25641, 0.28139)
(-0.256003, 0.280074) (-0.255597, 0.278763) (-0.254783, 0.276157)
(-0.253156, 0.271006) (-0.249903, 0.26094) (-0.243396, 0.241728)
(-0.242989, 0.240566) (-0.242582, 0.239409) (-0.241769, 0.237109)
(-0.240142, 0.232563) (-0.236888, 0.22368) (-0.230381, 0.206732)
(-0.229982, 0.205728) (-0.229584, 0.204727) (-0.228786, 0.202738)
(-0.227191, 0.198806) (-0.224001, 0.191123) (-0.217622, 0.176455)
(-0.217223, 0.175568) (-0.216824, 0.174685) (-0.216027, 0.172929)
(-0.214432, 0.169458) (-0.211242, 0.162678) (-0.204862, 0.149743)
(-0.20443, 0.148895) (-0.203997, 0.148051) (-0.203132, 0.146374)
(-0.201402, 0.143064) (-0.197942, 0.136612) (-0.191022, 0.124365)
(-0.190589, 0.123628) (-0.190157, 0.122894) (-0.189292, 0.121435)
(-0.187562, 0.118557) (-0.184101, 0.112952) (-0.177181, 0.102326)
(-0.176777, 0.101729) (-0.176374, 0.101135) (-0.175566, 0.0999548)
(-0.173952, 0.0976239) (-0.170723, 0.0930795) (-0.164264, 0.0844465)
(-0.16386, 0.0839266) (-0.163457, 0.0834089) (-0.162649, 0.0823804)
(-0.161035, 0.0803501) (-0.157806, 0.0763952) (-0.151347, 0.0688958)
(-0.15091, 0.0684071) (-0.150472, 0.0679208) (-0.149597, 0.0669553)
(-0.147847, 0.0650527) (-0.144348, 0.0613591) (-0.137349, 0.054404)
(-0.136911, 0.0539878) (-0.136474, 0.0535739) (-0.135599, 0.0527523)
(-0.133849, 0.0511346) (-0.130349, 0.0479989) (-0.12335, 0.0421143)
(-0.122921, 0.0417695) (-0.122491, 0.0414266) (-0.121632, 0.0407463)
(-0.119914, 0.0394075) (-0.116479, 0.0368163) (-0.109607, 0.0319683)
(-0.109177, 0.0316797) (-0.108748, 0.0313928) (-0.107889, 0.0308239)
(-0.106171, 0.0297057) (-0.102735, 0.027547) (-0.0958638, 0.0235302)
(-0.0954632, 0.0233081) (-0.0950626, 0.0230873) (-0.0942613, 0.0226495)
(-0.0926589, 0.0217893) (-0.089454, 0.0201299) (-0.0830443, 0.0170476)
(-0.0826436, 0.0168652) (-0.082243, 0.0166839) (-0.0814418, 0.0163249)
(-0.0798394, 0.015621) (-0.0766345, 0.0142683) (-0.0702247, 0.0117776)
(-0.0697903, 0.0116189) (-0.0693559, 0.0114614) (-0.0684871, 0.0111503)
(-0.0667495, 0.010543) (-0.0632743, 0.00938724) (-0.0628399, 0.00924821)
(-0.0624055, 0.00911037) (-0.0615366, 0.00883826) (-0.059799, 0.00830822)
(-0.0563238, 0.00730399) (-0.0558894, 0.00718363) (-0.055455, 0.0070644)
(-0.0545862, 0.00682933) (-0.0528486, 0.00637265) (-0.0493733, 0.00551235)
(-0.0489389, 0.00540972) (-0.0485045, 0.00530816) (-0.0476357, 0.00510827)
(-0.0458981, 0.00472128) (-0.0424228, 0.00399769) (-0.0420173, 0.00391757)
(-0.0416118, 0.00383834) (-0.0408007, 0.00368255) (-0.0391785, 0.00338157)
(-0.0359342, 0.00282143) (-0.0355286, 0.00275528) (-0.0351231, 0.00268998)
(-0.034312, 0.00256193) (-0.0326899, 0.00231594) (-0.0294455, 0.00186382)
(-0.02904, 0.001811) (-0.0286345, 0.00175899) (-0.0278234, 0.00165739)
(-0.0262012, 0.00146383) (-0.0257957, 0.00141743) (-0.0253901, 0.00137183)
(-0.0245791, 0.00128299) (-0.0229569, 0.00111474) (-0.0225514, 0.00107462)
(-0.0221458, 0.00103528) (-0.0213347, 0.000958904) (-0.0197126, 0.000815355)
(-0.019307, 0.00078137) (-0.0189015, 0.000748141) (-0.0180904, 0.000683944)
(-0.0164683, 0.000564529) (-0.0160707, 0.000537076) (-0.0156731, 0.000510334)
(-0.014878, 0.000458971) (-0.0144804, 0.000434347) (-0.0140829, 0.000410425)
(-0.0132878, 0.000364677) (-0.0128902, 0.000342849) (-0.0124926, 0.000321714)
(-0.0116975, 0.000281518) (-0.0101073, 0.000209364) (-0.00970969, 0.000193031)
(-0.00931213, 0.000177376) (-0.008517, 0.000148092) (-0.00811944, 0.000134459)
(-0.00772188, 0.000121497) (-0.00692675, 0) (-0.00652919, 0) (0.00619281, 0)
(0.00659037, 0) (0.0073855, 0.000107185) (0.00778306, 0.000118923)
(0.00818062, 0.000131259) (0.00897575, 0.000157717) (0.00940711, 0.000173064)
(0.00983846, 0.000189106) (0.0107012, 0.000223269) (0.0111325, 0.000241385)
(0.0115639, 0.000260189) (0.0124266, 0.000299848) (0.012858, 0.0003207)
(0.0132893, 0.000342231) (0.014152, 0.00038732) (0.0158775, 0.00048555)
(0.0163088, 0.000511774) (0.0167402, 0.000538659) (0.0176029, 0.000594408)
(0.0193283, 0.000713761) (0.0227792, 0.000983445) (0.0232105, 0.00102002)
(0.0236419, 0.00105723) (0.0245046, 0.00113352) (0.02623, 0.00129359)
(0.0296809, 0.00164322) (0.0301123, 0.00168966) (0.0305436, 0.00173669)
(0.0314063, 0.00183254) (0.0331318, 0.00203138) (0.0365826, 0.00245712)
(0.0369851, 0.00250918) (0.0373876, 0.00256174) (0.0381926, 0.00266835)
(0.0398026, 0.00288747) (0.0430225, 0.00334903) (0.0494624, 0.004363)
(0.0498649, 0.0044303) (0.0502674, 0.00449805) (0.0510724, 0.00463491)
(0.0526824, 0.00491404) (0.0559023, 0.00549357) (0.0623422, 0.00673557)
(0.0627785, 0.00682361) (0.0632148, 0.00691215) (0.0640873, 0.00709067)
(0.0658325, 0.0074535) (0.0693228, 0.00820195) (0.0763034, 0.00978743)
(0.0902646, 0.0132931) (0.0906929, 0.0134074) (0.0911213, 0.013522)
(0.0919779, 0.0137525) (0.0936911, 0.0142179) (0.0971176, 0.0151669)
(0.103971, 0.0171352) (0.117676, 0.0213374) (0.143241, 0.0300264)
(0.170968, 0.0405232) (0.196849, 0.0511575) (0.224891, 0.0634337)
(0.252424, 0.0761154) (0.278109, 0.0884096) (0.305957, 0.102156)
(0.331958, 0.115313) (0.357448, 0.128463) (0.3851, 0.142961)
(0.410906, 0.156671) (0.438875, 0.171694) (0.466332, 0.186582)
(0.491943, 0.200574) (0.519716, 0.215848) (0.545643, 0.23019)
(0.571058, 0.244324) (0.598637, 0.259741) (0.624368, 0.274201)
(0.652262, 0.289961) (0.678308, 0.304764) (0.703844, 0.319365)
(0.731543, 0.335313) (0.757395, 0.350313) (0.78541, 0.36671)
(0.812913, 0.382972) (0.83857, 0.398308) (0.86639, 0.415142)
(0.892362, 0.431074) (0.917823, 0.446922) (0.945448, 0.464403)
(0.971225, 0.481016) (0.971675, 0.481309) (0.972124, 0.481601)
(0.973023, 0.482187) (0.974822, 0.483359) (0.978419, 0.485709)
(0.985612, 0.490428) (0.986062, 0.490724) (0.986512, 0.49102)
(0.987411, 0.491612) (0.989209, 0.492797) (0.992806, 0.495174)
(0.993256, 0.495471) (0.993705, 0.495769) (0.994605, 0.496364)
(0.996403, 0.497557) (0.996853, 0.497855) (0.997302, 0.498154)
(0.998202, 0.498751) (0.998651, 0.49905) (0.999101, 0.499349)
(0.99955, 0.499648) (1., 0.499947)};
\end{tikzpicture}
\caption{A graph of the function $f(\delta)$ in \eqref{asymptoticseries} that controls the dimension of $\hat S$.\label{fig:one}}
\end{figure}
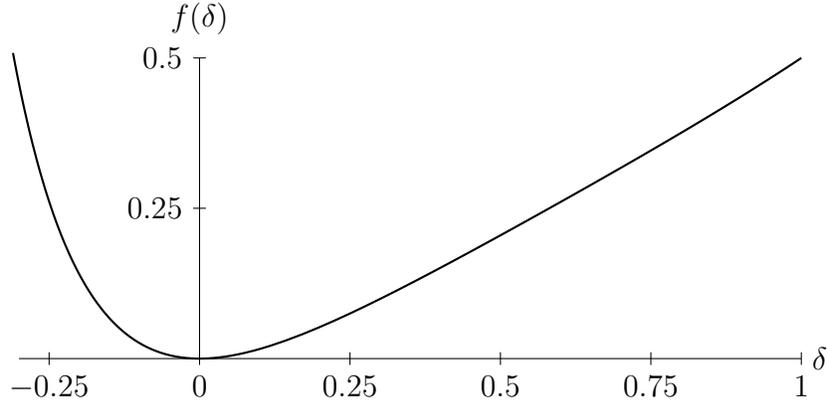

An alternative approach is to use the Padé-conformal method (see \cite{Beccaria:2021vuc} and references therein). In this case, the approximants up to the order that we computed give $f(1) \approx 0.500...\,$, where the given digits are completely stable. It is tempting to conclude that $f(1)=1/2$, and we use this value in what follows. See a graph of the function in Figure~\ref{fig:one}.

With relatively little further effort we can compute the dimensions of $\hphi$ and $\partial_\perp\hphi$. The resulting expressions are easier and match the expansion of rational functions \eqref{gammaphi}, \eqref{delphiresum}, which we resum to
\begin{align}
\label{dimphiint}
 \hat{\Delta}_{\hphi} &= 1+\delta+\frac{\veps}{2}-\frac{N+2}{N+8}\, \frac{\veps}{1+\delta}+ \text{O}(\veps^2)\,,
\\
\label{dimdelphi}
 \hat{\Delta}_{\del_\perp \hphi}&= 2-\frac{\veps}{2}-\frac{N+2}{N+8}\frac{\veps \, \delta}{(2-\delta)(1+\delta)}+ \text{O}(\veps^2)\,.
\end{align}

We are now ready to discuss what happens for different values of $\delta$. One obvious case is to set $\delta=1-\veps$, where the defect becomes an interface. Collecting the results for the observables that we computed, we find using $f(1-\veps) \approx f(1) + \text{O}(\veps)\approx\frac{1}{2}+\text{O}(\veps)$
\begin{equation}\label{interfaceresult}
\begin{aligned}
 \hat{\Delta}_{\hphi}\big|_{3-\veps} &= 2- \frac{N+5}{N+8}\lsp\veps +\text{O}(\veps^2) \,, \ &
 \hD_{\hat{S}} \big|_{3-\veps}&=4-\frac{3}{2}\frac{N+6}{N+8}\lsp\veps +\text{O}(\veps^2)\,, \\
 \hat{\Delta}_{\del_\perp \hphi}\big|_{3-\veps} &=2- \frac{N+5}{N+8}\lsp\veps+\text{O}(\veps^2) \,, \
 &\hat{\Delta}_{\hat{D}}\big|_{3-\veps}&= 4-\veps \,,
\end{aligned}
\end{equation}
where $\hat{D} \sim \hphi_a \del_\perp \hphi_a$ is the displacement operator, whose dimension is protected. Interestingly, the dimensions of $\hphi$ and $\del_\perp \hphi$ coincide at first non-trivial order in $\veps$, as they do in the free theory. Additionally, they both agree with the value of the boundary operator of the $O(N)$ model with Dirichlet boundary conditions (known also as the ordinary fixed point) \cite{Diehl:1981jgg, Diehl:1994pr, Diehl:1998mh}. Indeed, it was conjectured in~\cite{Krishnan:2023cff} that the defect deformation that we study should flow in $d=3$ and $p=2$ (so for $\veps=1$ and $\delta=0$) to two disconnected copies of the ordinary fixed point in the IR. While this feature persists for $\veps\neq1$ and $p=d-1$, according to \eqref{interfaceresult} the dimensions of $\hat S$ and $\hat D$ do not match at order $\veps$ with $\hat{S}$ slightly lighter. This shows that the interface CFT limit of our transdimensional defect is not equivalent to two copies of the $O(N)$ model with Dirichlet boundary conditions, since in the latter case the lightest singlet defect operator is the displacement. It could still be that summing higher orders will lead to an identity at $\veps=1$ and this is somewhat supported by a two-loop numerical extrapolation of the surface defect at the beginning of Appendix \ref{sec:deg}.

Taking $\delta=-1$ to construct an $O(N)$ symmetric line defect is unsuccessful as all of \eqref{deltaphi2}, \eqref{dimphiint} and \eqref{dimdelphi} diverge. This may be cured at higher orders in $\veps$, but recall from Section~\ref{sec:free} that in the free theory there is also no symmetry-preserving line defect at $\veps=1$.

Alternatively, for $\delta=1$, the defect becomes three-dimensional. In this case, defect correlators immediately yield an example of a non-local CFT in three dimensions. Indeed, by looking at the sector of operators that are uncharged under the $SO(d-p)$ rotational symmetry around the defect, which is a global symmetry from the point of view of defect operators, we find a CFT whose light spectrum is
\begin{equation}\label{3d1}
 \begin{split}
\hat{\Delta}_{\hphi}\big|_{p=3} &= 2+ \frac{3}{N+8}\lsp\veps +\text{O}(\veps^2)\,,
\qquad
\hD_{\hat{S}} \big|_{p=3}=4+\frac{N+14}{2(N+8)}\lsp\veps+\text{O}(\veps^2)\,.
 \end{split}
\end{equation}
Additionally, there are also sectors that are charged under $SO(d-p)$, where $d-p$ is generically non-integer, as investigated already in \cite{Binder:2019zqc}. For instance, we have
\begin{equation}\label{3d2}
\hat{\Delta}_{\del_\perp \hphi}\big|_{p=3} =2- \frac{N+5}{N+8}\lsp\veps +\text{O}(\veps^2) \,,
\qquad
\hat{\Delta}_{\hat{D}}\big|_{p=3} = 4\,.
\end{equation}
These charged operators might be evanescent\,\cite{Hogervorst:2015akt} for $\veps=1$. In fact, their correlators are necessarily proportional to contractions of projectors. For example, $\delta^{ij}\langle (\del_i \phi )( \del_j \phi )( \phi^2 )\rangle \propto 1-\veps$, which is vanishing for $\veps=1$.

For $\veps \rightarrow 0$ and $\delta\to1$ we recover the free field results in Section~\ref{sec:free}, namely a three-dimensional GFF with $\smash{\hD_{\hphi}}=2$. In the free theory, tuning $\veps\to1$ leads to a GFF with $\hD_\hphi=5/2$. In the interacting case here, turning on $\veps$ provides instead a continuous deformation of the three-dimensional GFF with $\hD_{\hphi}=2$ into an interacting CFT. There is a vast literature on deformations of GFF, which can be obtained holographically~\cite{Bertan:2018afl} or through RG flows~\cite{Fisher:1972zz, Paulos:2015jfa, Chai:2021arp}, yet our construction appears to be different. For example, in our deformation the dimension $\hD_{\hphi}$ depends non-trivially on $\veps$ as seen in \eqref{dimphiint}. This clearly illustrates the point that transdimensional defects give new ways to define CFTs.

\section{Large \texorpdfstring{$\boldsymbol{N}$}{N} analysis for
\texorpdfstring{$\boldsymbol{4<d<6}$}{4<d<6}}
\label{sec:largeN}
The large $N$ limit of the interacting $O(N)$ model allows to treat the theory analytically in the space dimension $d$. (For reviews see~\cite{Moshe:2003xn, Henriksson:2022rnm}.) In $d=4+\veps$ for $\veps>0$, the $O(N)$ model becomes an ultraviolet (UV) fixed point and it is natural to look for an alternative theory in which it is found as an IR fixed point. Such a theory would serve as the ``UV completion'' of the $O(N)$ model in $d=5$. This question was discussed in great detail in \cite{Fei:2014yja, Fei:2014xta, Gracey:2015tta, Antipin:2021jiw}, which found evidence that an $O(N)$ theory defined in $d=6-\veps$ may provide the desired UV completion, at least for sufficiently large $N$. The CFT in $d=5$ has been studied also through bootstrap methods \cite{Chester:2014gqa}. Generalisations to $d>6$ were considered in \cite{Stergiou:2015roa, Gracey:2015xmw, Osborn:2016bev}.

The bulk action
\begin{equation}\label{largeNaction}
 S=\int d^{\lsp d} x \, \Big( \tfrac12\lsp \partial^\mu \phi_a\lsp\partial_\mu\phi_a +\frac{1}{2\sqrt{N}}\sigma \phi_a \phi_a \Big)\,,
\end{equation}
where $\Delta_{\phi}=\frac{d-2}{2}+\text{O}(N^{-1})$ and $\Delta_\sigma=2+\text{O}(N^{-1})$. Co-dimension one defects for the large $N$ vector model in $d=3$ were studied in \cite{Krishnan:2023cff}. For $4<d<6$ one can look at surface defect with the action \cite{Trepanier:2023tvb, Giombi:2023dqs}
\begin{equation}\label{defectinteractionlargeN2}
S_D= \int d^{\lsp p} \tau \,( g_0 \,\phi_1(\tau) + h_0 \, \sigma(\tau))
\end{equation}
and $p=2$. It was shown there that, at leading order in $\veps$, the symmetry preserving defect (with $g=0$) in the $d=6-\veps$ theory matches the symmetry-preserving surface defect of the $d=4+\veps$ theory.

The symmetry-breaking defect was also studied in~\cite{Trepanier:2023tvb} in $d=6-\veps$. To go beyond that requires transdimensional defects, as $\phi_1$ is weakly relevant around $p=\frac{d-2}{2}$, which interpolates between a line defect in $d=4$ and a surface defect in $d=6$. Taking $p=\frac{d-2}{2}+\delta$ with small $\delta$ serving as a regulator, it is possible to use perturbation theory to compute observables at large $N$ for all $d$. Away from $d\simeq6$ we can consistently set $h_0=0$, as $\sigma$ is irrelevant, but close to $d=6$ it cannot be ignored and one returns to the $\veps$ expansion of~\cite{Trepanier:2023tvb}, which could be generalised to allow corrections in $\delta$.

If we take $\delta=2-d/2$ or $\delta=3-d/2$, the defect again becomes a line or a surface for any value of $d$ and requires high order analysis in $\delta$, which we do not pursue here. Instead we do the large $N$ analysis for small $\delta$ away from $d=6$.

To renormalise the defect coupling $g_0$, we compute the order parameter $\langle \phi_1(x) \rangle$ as usual. At lowest non-trivial order, this is given by the diagrams
\begin{equation}\label{diagram1largeN}
\begin{tikzpicture}[scale=0.5,baseline=(vert_cent.base)]
	 \draw[very thick,black] (-1.5,0)--(1.5,0);
	 \draw[thick, dashed] (0,3)--(0,0);
 \fill[black] (0,0) circle (4pt);
	 \node[above] at (0,3) {$\phi_1 $};
 \node[inner sep=0pt,outer sep=0pt] (vert_cent) at (0,1.75) {$\phantom{\cdot}$};
\end{tikzpicture}
\qquad
\begin{tikzpicture}[scale=0.5,baseline=(vert_cent.base)]
	 \draw[very thick,black] (-2,0)--(2,0);
	 \draw[thick, dashed] (0,3)--(0,2);
 \draw[thick, dashed] (0,2)--(-4/3,0);
 \draw[very thick, dotted] (0,2)--(2/3,1);
 \draw[thick, dashed] (2/3,1)--(4/3,0);
 \draw[thick, dashed] (2/3,1)--(0,0);
 \filldraw[color=black,fill=white,line width=0.35mm] (0,2) circle (4pt);
 \fill[black] (-4/3,0) circle (4pt);
 \fill[black] (4/3,0) circle (4pt);
 \fill[black] (0,0) circle (4pt);
 \filldraw[color=black,fill=white,line width=0.35mm] (2/3,1) circle (4pt);
	 \node[above] at (0,3) {$\phi_1$};
 \node[inner sep=0pt,outer sep=0pt] (vert_cent) at (0,1.75) {$\phantom{\cdot}$};
\end{tikzpicture}
\end{equation}
The dashed line is the $\phi_1$ propagator as in \eqref{eq:freeprop}, and the dotted line represents the propagator for the field $\sigma$
\begin{equation}
 \langle \sigma(x) \sigma(y) \rangle= \frac{\mathcal{N}_\sigma^{\,2}}{|x-y|^4} + \text{O}(N^{-1})\,, \qquad \mathcal{N}_\sigma^{\,2}= \frac{2^{d+2}\lsp\Gamma\! \left( \frac{d-1}{2}\right)\sin\!\left( \frac{\pi d}{2} \right)}{\pi^{3/2}\lsp\Gamma \!\left( \frac{d}{2}-2\right)}\,.
\end{equation}

Setting $g_0=\mu^{\delta}Z_g g$, we find
\begin{equation}\label{ZgN}
 Z_g = 1 + \frac{ c \, g^2}{N \lsp \delta}+\text{O}(N^{-2})\,,
\qquad
c=\frac{2^{d-3} \lsp\Gamma\!\left(\frac{d-1}{2}\right) \sin \!\left(\frac{\pi d}{4}\right)}{\pi ^{3/2}\lsp \Gamma\!\lf( \frac{d}{2}\rg)}\,.
\end{equation}
Note that at this order we don't need to take into account the wavefunction renormalisation of $\phi_1$, since it is subleading in $N$. The beta function to order $1/N$ is
\begin{equation}
 \beta_g= - \delta \lsp g + \frac{2 c \lsp g^3}{N}+ \text{O}(N^{-2})\,,
\end{equation}
and admits a fixed point for
\begin{equation}
 g^2_* = \frac{N \delta}{2 c}+\text{O}(\delta^{\lsp 2})\,.
\end{equation}
At this fixed point,
\begin{equation}\label{deltaphiN}
 \hD_{\hphi_1}= \frac{d}{2}-1+\delta+\frac{\del \beta_g}{\del g}\Big|_{g=g_*}\!\!
 =\frac{d}{2}-1+3\delta + \text{O}(N^{-1})\,.
\end{equation}

At higher orders in $\delta$ one needs to include all the contributions from tree diagrams generalising the one on the right of \eqref{diagram1largeN}. They contribute terms like $g^{2k+1}/N^k$ to $\beta_g$ to obtain a reliable result. The coefficient $c$ \eqref{ZgN} is negative for $4<d<6$, hence the fixed point value of the coupling is real when $\delta$ is negative and the fixed point is purely imaginary when $\delta$ is positive. In particular, for a line defect we need to take $\delta=(4-d)/2<0$, so we expect a real fixed point. On the other hand, for a surface defect we need to take $\delta=(6-d)/2>0$, which should give an imaginary fixed point.

We can check the results of this model against those obtained from the $\veps$ expansion. For $d=4+\veps$ and $\delta=-\veps/2$, \eqref{deltaphiN} gives $\hD_{\hphi_1}=1-\veps+\text{O}(N^{-1})$. This result is consistent with that derived for the symmetry-breaking line defect in $d=4+\veps$ \cite{Cuomo:2021kfm}.

Evaluating the one-point function of $\phi_1$ at the fixed point yields
\begin{equation}\label{largeNonepoint}
 \langle \phi_1 (x_\perp,x_\parallel) \rangle\big|_{g=g*}= \frac{\Nm_\phi\, a_\phi}{|\xp|^{\Delta_\phi}}\,,
\qquad
a_\phi^2=\frac{\Gamma\!\left(\frac{d-2}{4}\right)^2 \Gamma\!\lf(\frac{d}{2}\rg)}
 {8\sin\!\left(\frac{\pi d}{4}\right) \Gamma (d-2)} \, \delta\lsp N + \text{O}(N^0)\,,
\end{equation}
with $\Nm_\phi$ given by \eqref{eq:freeprop}. Expanding for $d=4+\veps$ and $\delta=-\veps/2$ we find
\begin{equation}
 a_\phi^2 \sim \frac{N}{4}-\frac{N \veps}{8} \lf( 1+ \log 4 \rg) +
 \text{O}(\veps^2)\,,
\end{equation}
which agrees with the large $N$ behaviour of the result obtained in \cite{Allais:2014fqa, Cuomo:2021kfm} after a change in the sign of $\veps$.

As mentioned, this solution is not valid around $d=6-\veps$, as the beta function of $h$ contains the classical term $(3-\frac{d}{2}-\delta)h$. For $3-\frac{d}{2}$ close to zero, even if one tries to fine-tune $h=0$, a non-trivial value of $h$ would be dynamically generated along the RG flow. Unfortunately, it is difficult to perform explicit calculations analytic in $d$ with $h_0\neq0$, due to the dimension of $\sigma$ being close to 2 for all $d$. Of course one can perform such calculations for $d$ close to 6, similarly to what we do in $d=4-\veps$ in Section~\ref{sec:interacting} above. Nevertheless, some observables can still be computed thanks to the equation of motion. Indeed, at any fixed point we expect that in addition to a non-zero $a_\phi$ as in \eqref{largeNonepoint} we have
\begin{equation}\label{largeNsolution}
\langle \sigma(\xp,x_\parallel)\rangle= \frac{\mathcal{N}_\sigma \,a_\sigma}{|\xp|^2}
\end{equation}
for some $a_\sigma$. Moreover, at leading order in $1/N$, we also expect that $\langle \sigma \phi_1 \rangle \sim \langle \sigma \rangle \langle \phi_1 \rangle$. Then the expectation value of the equation of motion of \eqref{largeNaction} yields
\begin{equation}\label{onepteom}
\left( -\Box + \frac{\mathcal{N}_\sigma \, a_\sigma}{\sqrt{N}|\xp|^2}\right) \frac{1}{|\xp|^{\Delta_\phi}}
= \frac{1}{|\xp|^{\Delta_\phi+2}}\lf( \Delta_\phi \lf( \Delta_\phi+2+p-d\rg)-\frac{\mathcal{N}_\sigma \, a_\sigma}{\sqrt{N}} \rg)=0\,.
\end{equation}
For $p=d/2-1+\delta$ the solution is
\begin{equation}\label{eq:asigmad}
 a_\sigma = \frac{1}{\mathcal{N_\sigma}} \frac{d-2}{2}\sqrt{N} \delta +\text{O}(N^{-1/2})\,.
\end{equation}
We can now expand for $d=6-\veps$ and $\delta=\veps/2$ yields
\begin{equation}
 a_\sigma \sim \frac{\sqrt{N \veps} }{4 \sqrt{6}}+ \text{O} ( \veps^{3/2})\,,
\end{equation}
which agrees with the large $N$ behaviour of the result for the symmetry-breaking surface defect of~\cite{Trepanier:2023tvb}.\footnote{There is a typo in ~\cite[Eq.\ (4.14)]{Trepanier:2023tvb}; at leading order the correct behaviour in $N$ is $h_*=-\sqrt{\pi N\veps}/4\sqrt{6}$.} In $d=4+\veps$ the expansion of \eqref{eq:asigmad} also agrees with the results of \cite{Cuomo:2021kfm}.

\section{Discussion}
\label{sec:discuss}

This note introduced transdimensional defects whose dimension is a free continuous parameter and demonstrated their utility in several examples. This includes defects with dimensions $2\leq p\leq3$ in Euclidean space of dimension $3\leq d\leq4$, which are studied in the $\veps$ expansion, and defects of dimension $1\leq p\leq2$ in $4\leq d\leq6$ studied in the large-$N$ expansion.

For the interacting $O(N)$ model in $d=4-\veps$ we were able to perform very detailed calculations on a symmetry preserving defect of dimension $p=2+\delta$. Working at leading order in $\veps$ and to all orders in $\delta$, we were able to extrapolate to an interface with $p=3-\veps$ with low-lying spectrum presented in \eqref{interfaceresult}. This interface is different from those studied in~\cite{Harribey:2023xyv, Harribey:2024gjn}, which all break the global symmetry. In~\cite{Krishnan:2023cff} it was conjectured that such a defect with $\veps=1$ and $\delta=0$ is an impregnable interface with Dirichlet boundary conditions~\cite{Diehl:1981jgg, Diehl:1994pr, Diehl:1998mh}. The fact that $\hphi$ and $\partial_\perp\hphi$ have equal dimensions is similar to that system, but it deviates from it since the dimensions of $\hat S$ and $\hat D$ are not equal, and neither can $\hat S$ be interpreted as the product of defect primaries belonging to disjoint copies of a boundary theory, as in \cite{Diatlyk:2024ngd}. This is therefore a completely new interface in $d=4-\veps$.

Another special value is $\delta=1$, giving a three-dimensional defect in $d=4-\veps$. Focusing solely on defect operators, we treat this system as a non-local 3d CFT. We present the anomalous dimensions in~(\ref{3d1}, \ref{3d2}) and those do not match any theory known to us. Setting $\veps=0$ returns us to the free theory in $d=4$ and with the restriction of operators to $p=3$, this behaves like a theory of generalised free fields (GFF) with $\hD_\hphi=2$. Remaining in the free theory we can tune $\veps$ freely, as explained in Section~\ref{sec:free}, and for $\veps=\delta=1$ the defect is space-filling and we find another three-dimensional theory, again GFF but now with $\hD_\hphi=5/2$. The defect theory at other values of $\veps$ interpolates between these two values. Note that this is markedly different from starting with a quadratic deformation of the free theory directly in $d=p=3$. In that case this is a mass term and the IR is the empty theory.

It would be very interesting to extend the results in the interacting theory to higher orders in $\veps$ beyond those presented in Sections~\ref{sec:interacting}. As explained at the top of Section~\ref{sec:interacting}, going to high order in $h$ (or $\delta$) is relatively easy at a fixed order in $\veps$, so this is not beyond the realm of possibilities.

While all our calculations can be immediately applied also to symmetry breaking defects, we explored them in less detail. In Appendix~\ref{sec:deg} we extend the results of~\cite{Trepanier:2023tvb, Raviv-Moshe:2023yvq, Giombi:2023dqs} for both the symmetry preserving and symmetry breaking surface defects to higher order. In particular we show how some degenerate solutions noted in~\cite{Trepanier:2023tvb} are lifted at order $\veps^2$ with square root branch cuts in the defect coupling $h$.

The second case we study is the $O(N)$ model in $4\leq d\leq6$, which we approach using large $N$ techniques. It was shown in~\cite{Trepanier:2023tvb, Giombi:2023dqs} that the properties of the symmetry preserving surface defect interpolate nicely across dimensions. On the other hand, the natural defects breaking $O(N)\to O(N-1)$ are a line close to $d=4$ and a surface close to $d=6$. As we show in Section~\ref{sec:largeN}, this is a perfect setting for transdimensional defects and as long as $d=2p+2$ there is a good symmetry breaking defect of dimension $p$. The dimensions and bulk one-point function coefficients we obtain at large $N$ analytically in $d$ match with the expected answers when expanded around $d=4$ and $d=6$.

As demonstrated, transdimensional defects provide a practical tool to deal with non-weakly relevant defect deformations. This is an alternative method to massive approaches~\cite{Parisi:1980gya, Diehl:1998mh} and to finding solutions to the equations of motion with non-trivial defect profile~\cite{Shpot:2019iwk,Dey:2020lwp, Metlitski:2020cqy}. The former involves complicated diagrammatic calculations, since the massless limit is taken only at the end, while the latter works particularly well in the co-dimension one case, but it is difficult to carry out for other values of the defect dimension.

This note is meant as a proof of principle and we leave many generalisations to the curious reader. One can further develop the examples studied here to higher orders in $\veps$ and/or $1/N$, and calculate more anomalous dimensions and look for structure constants or higher point functions. Natural further examples are deformations away from $p=1$ lines or $p=3-\veps$ interfaces. Other observables involving transdimensional defects are defect four-point functions, bulk two-point functions, and so on. To those observables one could also apply analytic and numerical bootstrap techniques, as in~\cite{Liendo:2012hy, Gliozzi:2015qsa, Bissi:2018mcq, Kaviraj:2018tfd, Mazac:2018biw, Dey:2020jlc, Liendo:2019jpu, Barrat:2022psm, Bianchi:2022ppi, Bianchi:2022sbz, Bianchi:2023gkk}.

Another natural extension is to define transdimensional defects for other bulk fixed points beyond the $O(N)$ model. The only requirement is that the model admit a perturbative defect that can be defined consistently in non-integer dimension. For example, scalar models with different bulk symmetries \cite{Pannell:2023pwz} and theories with fermionic degrees of freedom \cite{Giombi:2022vnz, Pannell:2023pwz, Barrat:2023ivo}.

We conclude with some comments and speculations about transdimensional defect conformal manifolds. In the free $O(N)$ model in 4d the operator $\phi_a$ is exactly marginal when integrated along a line, $\phi_a\phi_b$ along a surface and likewise $\phi^3$ operators along an interface, so the corresponding defect coupling $h$ is arbitrary. When going to $4-\varepsilon$, the dimension of these operators changes even in the free theory and they are no longer marginal on an integer dimensional defect. Allowing for transdimensional defects, we can choose $p=2-\veps$ where $\phi_a\phi_b$ is again exactly marginal.

The analogous story in the interacting theory is more subtle. To leading order in $\veps$, the dimension of the scalar $\hat{S}$ in \eqref{ST} is given in \eqref{deltaphi2}. To make the operator exactly marginal we need higher corrections in $h$, where from \eqref{interactingbeta}
\begin{equation}
\delta_*=-\frac{6}{N+8}\veps+h
-2h\veps\frac{N+2}{N+8}\Big(1-h+\frac{6-\zeta(3)}{4}h^2+\text{O}(h^3)\Big)\,.
\end{equation}
Thus, we can view transdimensional defects as allowing nearly marginal operators like $S$ to be exactly marginal by a choice of $\delta_*(h)$.

A similar statement for the bulk theory is to reverse \eqref{wilsonfisherfixedpoint} and define $\veps_*(\lambda)$, so the quartic interaction term in \eqref{ondefectaction} is exactly marginal when combined with the change of dimension. This is not a common (or particularly useful) viewpoint for the $\veps$ expansion, but it may be more pertinent in the case of transdimensional defects with symmetry breaking, generalising \cite{Drukker:2022pxk}. For example, in the case of the defect in Section~\ref{sec:largeN}, the choice of $\phi_1$ in \eqref{defectinteractionlargeN2} breaks $O(N)\to O(N-1)$ giving a conformal manifold $S^{N-1}$. In the free theory in integer dimension, $h_0$ is also a marginal parameter, extending the manifold to $\mathbb{R}^N$. A similar structure may arise in the interacting theory for $p=\frac{d}{2}-1+\text{O}(N^{-1})$ when $\hphi_1$, whose dimension is given by \eqref{deltaphiN}, is marginal. Similar considerations can apply for other symmetry-breaking transdimensional defects.

\ack{We are grateful to L.\ Bianchi and M.\ Tr\'epanier for useful discussions. EdS's research was partially supported by a short term scientific mission grant from the COST action CA22113 THEORY-CHALLENGES. ND would like the acknowledge the hospitality of CERN, DESY, EPFL, Perimeter Institute and the Simons Center for Geometry and Physics in the course of this work. ND's research is supported in part by the Science Technology \& Facilities Council (STFC) under the grants ST/P000258/1 and ST/X000753/1. AS thanks the CERN Department of Theoretical Physics for hospitality in the course of this work. AS is supported by the Royal Society under grant URF\textbackslash{}R1\textbackslash211417 and by STFC under grant ST/X000753/1.}

\begin{appendices}
\section{Feynman integrals}\label{appintegrals}
In this appendix we evaluate the integrals appearing in Sections~\ref{sec:free}, \ref{sec:interacting}.

We start with the diagrams contributing to the bulk one-point function of $\phi_a\phi_b$
\begin{equation}\label{diagram2}
\begin{tikzpicture}[baseline=(vert_cent.base),scale=0.5]
 \draw[very thick,black] (-1.5,0)--(1.5,0);
 \draw[thick, dashed] (0,3) to[out=-45,in=45](0,0);
 \draw[thick, dashed] (0,3) to[out=-135,in=135](0,0);
 \fill[black] (0,0) circle (4pt);
 \node[above] at (0,3) {$\phi_a\phi_b$};
 \node[inner sep=0pt,outer sep=0pt] (vert_cent) at (0,1.75) {$\phantom{\cdot}$};
\end{tikzpicture}
\qquad
\begin{tikzpicture}[baseline=(vert_cent.base),scale=0.5]
 \draw[very thick,black] (-2.5,0)--(2.5,0);
 \draw[thick, dashed] (0,3)--(1.5,0);
 \draw[thick, dashed] (0,3)--(-1.5,0);
 \draw[thick, dashed] (-1.5,0)to[out=60,in=120](1.5,0);
 \fill[black] (-1.5,0) circle (4pt);
 \fill[black] (1.5,0) circle (4pt);
 \node[above] at (0,3) {$\phi_a\phi_b$};
 \node[inner sep=0pt,outer sep=0pt] (vert_cent) at (0,1.75) {$\phantom{\cdot}$};
\end{tikzpicture}
\qquad
\begin{tikzpicture}[baseline=(vert_cent.base),scale=0.5]
 \draw[very thick,black] (-1.5,0)--(1.5,0);
 \draw[thick, dashed] (0,3) to[out=-20,in=20](0,1.5);
 \draw[thick, dashed] (0,3) to[out=-160,in=160](0,1.5);
 \draw[thick, dashed] (0,1.5) to[out=-20,in=20](0,0);
 \draw[thick, dashed] (0,1.5) to[out=-160,in=160](0,0);
 \fill[black] (0,0) circle (4pt);
 \filldraw[color=black,fill=white,line width=0.35mm] (0,1.5) circle (4pt);
 \node[above] at (0,3) {$\phi_a\phi_b$};
 \node[inner sep=0pt,outer sep=0pt] (vert_cent) at (0,1.75) {$\phantom{\cdot}$};
\end{tikzpicture}
\end{equation}
It is convenient to define
$(h_0)_{ab}=\mu^{\veps+\delta}(h_{ab}+ \mathcal{Z}_{h,ab})$, where
$\mathcal{Z}_{h,ab}$ is an additive counterterm, and proceed to renormalise
the coupling such that the bulk one-point function
$\cor{\phi_a\phi_b(\xp,0)}$ is finite, as in~\cite{Trepanier:2023tvb, Raviv-Moshe:2023yvq, Giombi:2023dqs} but with non-zero $\delta$. Note that now the renormalised bulk one-point function also includes the wavefunction renormalisation factors $Z_S^{-1}$ and $Z_T^{-1}$ of the bulk operators $S$ and $T_{ab}$.

The diagrams in \eqref{diagram2} are regulated with both $\veps$ and $\delta$
and can be renormalised with
\begin{equation}\label{countertermhab}
 \mathcal{Z}_{h,ab}=\frac{2\lambda h_{ab}}{\veps}+\frac{\lambda \delta_{ab} h_{cc}}{\veps}+\frac{h_{ac}h_{cb}}{\veps+\delta}+\text{O}(h^3,h^2\lambda,h \lambda^2)\,.
\end{equation}
To specify the renormalisation scheme at higher orders, note that the Feynman integrals admit the unique decomposition
\begin{equation}\label{divergence}
 I_L(\veps,\delta)=\sum_{k=0}^L \prod_{i=1}^k
\frac{D_L}{\veps+\alpha_{k,i}\lsp \delta}+ g(\veps,\delta)\,,
\end{equation}
with $D_L$ and $\alpha_{k,i}$ real constants and $g(\veps,\delta)$ regular near $\veps=\delta=0$. The reason for this form is that the integrals depend on $\delta$ only through the defect dimension $p$, and we also know that by taking $\delta \rightarrow0$ while keeping $\veps$ finite we should retrieve the divergent integrals of the surface defect. From \eqref{countertermhab} we obtain the beta function
\begin{equation}\label{betafunctionhab}
 \beta_{ab}=-(\veps+\delta)h_{ab}+\lf(\delta_{ab}h_{cc}+2h_{ab}\rg) \lambda+h_{ac}h_{cb}+\text{O}(h^2\lambda,h \lambda^2)\,,
\end{equation}
where, besides the rescaling $\lambda\to16\pi^2\lambda$, we have also rescaled $h\to 2\pi h$.

We proceed now to look at Feynman diagrams of higher order in $h$, which allows to write expressions that are exact in $\delta$, at leading order in $\veps$. The corrections to the middle diagram in \eqref{diagram2} with one bulk vertex and a minimal number of defect vertices are
\begin{equation}\label{diagram4}
\begin{tikzpicture}[baseline=(vert_cent.base),scale=0.5]
	 \draw[very thick,black] (-3,0)--(3,0);
	 \draw[thick, dashed] (0,3)--(1.5,0);
	 \draw[thick, dashed] (0,3)--(-1,1.5);
 \draw[thick, dashed] (1.5,0)to[out=150,in=-30](-1,1.5);
 \draw[thick, dashed] (-1.5,0)to[out=120,in=210](-1,1.5);
 \draw[thick, dashed] (-1.5,0)to[out=45,in=285](-1,1.5);
 \fill[black] (-1.5,0) circle (4pt);
 \fill[black] (1.5,0) circle (4pt);
 \filldraw[color=black,fill=white,line width=0.35mm] (-1,1.5) circle (4pt);
	 \node[above] at (0,3) {$\phi_a\phi_b$};
 \node[inner sep=0pt,outer sep=0pt] (vert_cent) at (0,1.75) {$\phantom{\cdot}$};
	\end{tikzpicture}
	\hspace{1 cm}
 \begin{tikzpicture}[baseline=(vert_cent.base),scale=0.5]
	 \draw[very thick,black] (-3,0)--(3,0);
	 \draw[thick, dashed] (0,3)--(1.5,0);
	 \draw[thick, dashed] (0,3)--(-1.5,0);
 \draw[thick, dashed] (-1.5,0)to[out=60,in=180](0,1.2);
 \draw[thick, dashed] (0,1.2)to[out=0,in=120](1.5,0);
 \draw[thick, dashed] (0,1.2)to[out=-45,in=45](0,0);
 \draw[thick, dashed] (0,1.2)to[out=215,in=135](0,0);
 \fill[black] (-1.5,0) circle (4pt);
 \fill[black] (0,0) circle (4pt);
 \filldraw[color=black,fill=white,line width=0.35mm] (0,1.2) circle (4pt);
 \fill[black] (1.5,0) circle (4pt);
	 \node[above] at (0,3) {$\phi_a \phi_b$};
 \node[inner sep=0pt,outer sep=0pt] (vert_cent) at (0,1.75) {$\phantom{\cdot}$};
	\end{tikzpicture}
\end{equation}
To simplify the expressions, we treat explicitly only for the $O(N)$ symmetric defect, so $h_{ab}=h\delta_{ab}$, but the same calculations can be used for the symmetry breaking defect.

For the first graph in \eqref{diagram4}, we take the bulk external point to be $x=(\xp,x_\parallel)=(\mathbf{1},0)$, where $\mathbf{1}$ is any unit vector normal to the defect, and find
(up to symmetry factors, couplings and indices)
\begin{equation}
\begin{aligned}
I(x_\perp)=
\raisebox{0mm}[10mm][0mm]%
{\begin{tikzpicture}[baseline=(vert_cent.base),scale=0.5]
\node[inner sep=0pt,outer sep=0pt] (vert_cent) at (0,1.6) {$\phantom{\cdot}$};
	 \draw[very thick,black] (-3,0)--(3,0);
	 \draw[thick, dashed] (0,3)--(1.5,0);
	 \draw[thick, dashed] (0,3)--(-1,1.5);
  \draw[thick, dashed] (1.5,0)to[out=150,in=-30](-1,1.5);
  \draw[thick, dashed] (-1.5,0)to[out=120,in=210](-1,1.5);
  \draw[thick, dashed] (-1.5,0)to[out=45,in=285](-1,1.5);
  \fill[black] (-1.5,0) circle (4pt);
  \fill[black] (1.5,0) circle (4pt);
  \filldraw[color=black,fill=white,line width=0.35mm] (-1,1.5) circle (4pt);
	 \node[above] at (0,3) {$\phi_a\phi_b$};
\end{tikzpicture}}
= \smash{\int} d^{\lsp 4-\veps}y \smash{\int} d^{\lsp 2+\delta }\tau_1 \smash{\int} d^{\lsp 2+\delta }\tau_2  \,& G\!\left(y-\tau_1 \right)^2 G\!\left(y-\tau_2 \right)
\\[-4pt]&\hspace{16pt}G\!\left(x-\tau_2 \right)G\!\left(x-y \right).
\end{aligned}
\end{equation}
The integral over $\tau_1$ is trivial. After that, we are left with
\begin{equation}
\begin{split}
I(x_\perp) &= \Nm_\phi^{\lsp 10} \frac{\pi ^{1+\frac{\delta}{2}} \lsp \Gamma\!
    \left(1-\veps-\frac{\delta}{2}\right)}{\Gamma (2-\veps)} \int
    d^{\lsp 2-\veps-\delta} \yp \int d^{\lsp 2+\delta} y_\parallel \int
    d^{\lsp 2+\delta }\tau_2 \\
    & \quad\ \frac{1}{|\yp|^{2-2\veps-\delta}\lf( |\yp|^2+|\tau_2-y_\parallel|^2\rg)^{1-\frac{\veps}{2}}\lf( 1+\tau_2^2\rg)^{1-\frac{\veps}{2}}\lf( |\yp-\mathbf{1}|^2+|y_\parallel|^2\rg)^{1-\frac{\veps}{2}}}\,,
     \end{split}
\end{equation}
where $\Nm_\phi^{\lsp 2}$ is defined in \eqref{eq:freeprop}.

To evaluate this integral, it is convenient to go to defect momentum space. In fact, the two defect integrals are a convolution of the last three propagators, thus they can be exchanged for a single integral over the parallel momentum of the product of the defect Fourier transform of such propagators. The defect Fourier transforms of the last three propagators can be represented using Schwinger parameters,
\begin{equation}
 \int\frac{d^{\lsp 2+\delta} y_\parallel}{\lf( 2\pi \rg)^{2+\delta} } \, \frac{e^{i p_\parallel \cdot y_\parallel} }{\lf(|\yp|^2+|y_\parallel|^2\rg)^{1-\frac{\veps}{2}}}= \frac{\pi^{1+\frac{\delta}{2} }}{\Gamma\!\lf( 1-\frac{\veps}{2}\rg)} \int_0^\infty du \, u^{-1-\frac{\veps+\delta}{2}} \exp\lf(-u |\yp|^2-\frac{|p_\parallel|^2}{4u}\rg).
\end{equation}
After having introduced three parameters $u_1$, $u_2$ and $u_3$, one can easily perform first the $\yp$ integration and then the $p_\parallel$ integration. We are left with
\begin{equation}
\begin{split}
I(x_\perp)&= \frac{\Gamma\!\left(\frac{1}{2}-\frac{\veps+\delta}{2}\right)
\Gamma\!\left(1-\veps-\frac{\delta}{2}\right)}
{2^{10-\veps-\delta}
\pi^{3-\frac{5}{2}(\veps+\delta)}
\sin\!\left(\frac{\pi \veps}{2}\right)
\Gamma\!\left(\frac{3}{2}-\frac{\veps}{2}\right)}
\int_0^\infty du_1\int_0^\infty du_2 \int_0^\infty du_3
\\&\hskip2cm
e^{-u_2-u_3}
\frac{(u_1 u_2 u_3 (u_1+u_2))^{-\veps/2}}
{(u_1 (u_2+u_3)+u_2 u_3)^{1+\delta/2}}
_1\tilde{F}_1\left(\frac{\veps}{2};1-\frac{\veps+\delta}{2};\frac{u_2^2}{u_1+u_2}\right),
\end{split}
\end{equation}
where $_1\tilde{F}_1$ is the regularised confluent hypergeometric function
\begin{equation}\label{confluenthyp}
 _1\tilde{F}_1(a;b;z)=\frac{1}{\Gamma(a)} \sum_{k=0}^\infty \frac{ \Gamma (a+k)}{k! \,\Gamma (b+k)}z^k \,.
\end{equation}
The prefactor outside the integral has a simple pole, whereas the integral over the parameter $u_1$ is still divergent in the limit $\veps\rightarrow 0$ and $\delta\rightarrow 0$ (the other integrals are finite). To extract the divergent terms we first expand the hypergometric and then perform the $u_1$ integral. Crucially, only the first term of the expansion \eqref{confluenthyp} contributes to the divergence of $I$, since all other terms are of order $\text{O}(\veps, \delta)$. Therefore, we find
\begin{equation}
 I= K(\veps,\delta) \int_0^\infty du_2 \int_0^\infty du_3 \, f(u_2, u_3, \veps,\delta) + \text{finite}\,,
\end{equation}
where now the integrals are convergent and the poles are explicit in $K(\veps,\delta)$ and $f(u_2, u_3, \veps,\delta)$. The integral is still hard to compute, since $f(u_2, u_3, \veps,\delta)$ contains Gauss's hypergeometric functions $_2F_1$. To extract the divergent behaviour in the form of \eqref{divergence}, we make the substitution $\delta \rightarrow \alpha \lsp\veps$ and then expand in powers of $\veps$, neglecting regular terms. The resulting integrals are then easily performed. At the end, exploiting the parameter $\alpha$, we can reconstruct the dependence on $\veps$ and $\delta$, finding
\begin{equation}\label{divergence1}
 I=\frac{1}{256\pi^6 \, \veps(2\veps+\delta)}+ \frac{1+\gamma_E+\log \pi }{256 \pi^6 \veps}+\frac{1}{128\pi^6(2\veps+\delta)}+\text{finite}\,.
\end{equation}

Alternatively, the integral may be computed as follows. The relevant divergence comes from the region where the bulk interaction point gets close to the defect (the divergence associated with the interaction point getting close to the external point is accounted for with $Z_{\phi^2}$). To compute the divergence, we integrate only over $|\yp|<\eta$ with $\eta$ small. The integral becomes considerably simpler because we can replace, e.g., $|\yp-\mathbf{1}| \rightarrow 1$. One needs to first do all other integrations, expanding in $\yp$ and keeping terms divergent neat $\yp \sim 0$. Then the final integral over $\yp$ reproduces \eqref{divergence1}.

The second diagram in \eqref{diagram4} is easier, but we first evaluate a sub-diagram which (up to symmetry factors, couplings and indices) is
\begin{align}\label{diagramApp}
I'(\tau)&=
\begin{tikzpicture}[baseline=(vert_cent.base),scale=0.5]
	 \draw[very thick,black] (-3,0)--(3,0);
 \draw[thick, dashed] (-1.5,0)to[out=90,in=180](0,1.5);
 \draw[thick, dashed] (0,1.5)to[out=0,in=90](1.5,0);
 \draw[thick, dashed] (0,1.5)to[out=-45,in=45](0,0);
 \draw[thick, dashed] (0,1.5)to[out=215,in=135](0,0);
 \fill[black] (0,0) circle (4pt);
 \filldraw[color=black,fill=white,line width=0.35mm] (0,1.5) circle (4pt);
 \node[below] at (-1.5,0) {$\hphi_a(0)$};
 \node[below] at (1.5,0) {$\hphi_a(\tau)$};
 \node[inner sep=0pt,outer sep=0pt] (vert_cent) at (0,0) {$\phantom{\cdot}$};
\end{tikzpicture}
=
\int d^{\lsp 4-\veps}x \int d^{\lsp 2+\delta}\tau_1 \, G\!\left(x-\tau_1 \right)^2 G\!\left(x\right) G\!\left(x-\tau\right)\nonumber\\
 &= \frac{\Nm_\phi^8}{\tau^{2-3\veps-\delta}}
 \int d^{\lsp 4-\veps}x \int\frac{d^{\lsp 2+\delta } \tau_1}{\lf(|\xp|^2+|x_{\parallel}|^2 \rg)^{1-\frac{\veps}{2}}\lf(|\xp|^2+|x_{\parallel}-\tau_1|^2 \rg)^{2-\veps}\lf(|\xp|^2+|x_{\parallel}-\mathbf{1}|^2 \rg)^{1-\frac{\veps}{2}}}\,.
\end{align}
This integral can be computed exactly by integrating over $\tau_1$ and then introducing a Feynman parameter for the first and the third propagators and yields
\begin{equation}\label{diagApp1}
I'(\tau)=\frac{1}{\tau^{2-3\veps-\delta}}
\frac{\Gamma\!\left(1-\frac{3\veps+\delta}{2}\right)}
{2^{8+\veps+\delta} \pi^{2-\frac{3\veps+\delta}{2}} \sin\!\left(\pi\frac{\veps}{2}\right)
\sin\!\left(\pi\frac{2\veps+\delta}{2}\right)
\Gamma\!\left(\frac{3-\veps}{2}\right)
\Gamma\!\left(1-\frac{\veps+\delta}{2}\right)
\Gamma\!\left(\frac{1}{2}+\frac{2\veps+\delta}{2}\right)}\,.
\end{equation}
From this we can find the pole structure in the form \eqref{divergence}
\begin{equation}
 I'(\tau)=\frac{1}{\tau^{2-3\veps-\delta}}\bigg(\frac{1}{64\pi^5 \, \veps \lf(\veps+\frac{1}{2}\delta \rg)}+ \frac{\gamma_E +\log \pi}{64 \pi^5 \veps}+\frac{2+\gamma_E +\log \pi}{128\pi^5\lf( \veps+\frac{1}{2}\delta\rg)}+\text{O}(\veps^0,\delta^0)\bigg)\,.
\end{equation}
Then the second diagram of \eqref{diagram4} follows easily from \eqref{diagApp1} by integrating over $\tau$.

We now turn to evaluating high orders in $\delta$. The effective defect-to-defect propagator \eqref{effproprecursion} is
\begin{equation}\label{effpropapp}
 \begin{tikzpicture}[scale=0.5, baseline=(vert_cent.base)]
 \node[inner sep=0pt,outer sep=0pt] (vert_cent) at (0,0) {$\phantom{\cdot}$};
 \draw[very thick,black] (-3.5,0)--(3.5,0);
 \draw[thick, dashed] (-3,0)to[out=90,in=180](-1.5,1.5)to[out=0,in=90](0,0);
 \draw[thick, dashed] (3,0)to[out=90,in=0](1.5,1.5)to[out=180,in=90](0,0);
 \draw[fill= white] (0,0) circle (15pt);
 \draw[fill= white, draw=black, line width=0.25mm, pattern=north east lines] (0,0) circle (15pt);
 \node[below] at (-3,0) {$0$};
 \node[below] at (3,0) {$\tau$};
 \end{tikzpicture}
 = \sum_{k=0}^\infty h_0^k \frac{a_k}{|\tau|^{2-(k+1)\veps-k \delta}} \,,
\qquad
a_k= \frac{(-1)^k\lsp\Gamma\!\left(\frac{\veps+\delta}{2}\right)^{k+1} \Gamma\!\left(1-\frac{k+1}{2}\veps-\frac{k}{2} \delta\right)}{2^{k+2} \pi^{2-\frac{k+1}{2}\veps-\frac{k}{2}\delta}\lsp\Gamma\!\left(\frac{k+1}{2}(\veps+\delta)\right)}\,.
\end{equation}
As before, we strip the diagrams of indices, which however can be easily restored.

This is then inserted into \eqref{bulkeffproprecursion}, but it is still difficult to solve in general the remaining integrals. However, for our purposes it is sufficient to compute them in two specific instances. The first one is the case where the two external points coincide, which is diagram (\ref{diagrams1sc}a). This gives an effective one-point function that reads
\begin{equation}\label{onepointallorderapp}
\begin{tikzpicture}[scale=0.5, baseline=7mm]
	 \draw[very thick,black] (-2,0)--(2,0);
	 \draw[thick, dashed] (0,3) to[out=-45,in=45](0,0);
	 \draw[thick, dashed] (0,3) to[out=-135,in=135](0,0);
 \draw[fill= white] (0,0) circle (15pt);
 \draw[fill= white, draw=black, line width=0.25mm, pattern=north east lines] (0,0) circle (15pt);
 \node[above] at (0,3) {$\phi^2(x)$};
\end{tikzpicture}
 \ = \ \sum_{k=0}^\infty h_0^{k+1} \frac{b_k}{|\xp|^{2-(k+2)\veps-(k+1) \delta}} \,,
\end{equation}
with
\begin{equation}\label{bkcoeff}
\begin{split}
 b_k= \frac{(-1)^k\Gamma\!\left(\frac{\veps+\delta}{2}\right)^k \Gamma\!\left(1-\frac{k}{2}\veps-\frac{k-1}{2} \delta\right) \Gamma \!\left(1-\frac{k+1}{2}\veps-\frac{k}{2} \delta\right) \Gamma\!\left( 1-\frac{k+2}{2}\veps-\frac{k+1}{2} \delta\right) }{2^{k+4-(k+1)\veps-k\delta}\pi^{\frac{3}{2}-\frac{k+2}{2}\veps-\frac{k+1}{2}\delta}\Gamma\!\left(1+\frac{\delta}{2}\right) \Gamma\!\left(\frac{3}{2}-\frac{k+1}{2}\veps-\frac{k}{2} \delta\right)} \,.
 \end{split}
\end{equation}
The renormalised one-point function in \eqref{freeonepoint} is obtained by substituting the renormalised coupling in \eqref{onepointallorderapp} and evaluating at the non-trivial fixed point of \eqref{eq:betahfree}. By expanding in $\veps$ and $\delta$ one can check that the series agrees with the result given in \eqref{freeonepoint}.

The last building block needed for renormalisation is the limit of \eqref{bulkeffproprecursion} with one point close to the defect and one far away. Including an extra propagator between these points, this is
\begin{equation}\label{cd}
\int d^{\lsp 2+\delta} x_\parallel \Bigg(\,
\begin{tikzpicture}[scale=0.35, baseline= 0.5 cm]
	 \draw[very thick,black] (-2,0)--(3,0);
	 \draw[thick, dashed] (0.5,3.5)--(1,0);
	 \draw[thick, dashed] (-0.5,0.7)--(0.5,3.5);
 \draw[thick, dashed] (1,0)--(-0.5,0.7);
 \draw[fill= white] (1,0) circle (20pt);
 \draw[fill= white, draw=black, line width=0.25mm, pattern=north east lines] (1,0) circle (20pt);
	 \node[right] at (0.7,3.2) {$y$};
 \node[left] at (-.5,1) {$x$};
\end{tikzpicture}
\, \Bigg)
 =
 \sum_{k=0}^\infty \frac{h_0^{k+1}}{|y_\perp|^{2-(k+4)\veps-(k+3)\delta}} \Bigg( c_k + d_k \left( \frac{|x_\perp|}{|y_\perp|}\right)^{\veps+\delta} \!\!\!\! + \text{O}\!\left( \frac{|x_\perp|}{|y_\perp|} \right)\!\!\Bigg) \,,
\end{equation}
where $x=(x_\parallel,x_\perp)$, $y=(y_\parallel,y_\perp)$, and
\begin{equation}\label{ckcoeff}
 c_k= \frac{ (-1)^k\Gamma\! \left(\frac{\veps+\delta}{2}\right)^{k+2} \Gamma\!\left( 1-\frac{k+2}{2}\veps-\frac{k+1}{2} \delta \right) \Gamma\!\left(1-\frac{k+3}{2}\veps-\frac{k+2}{2} \delta \right) \Gamma\!\left( 1-\frac{k+4}{2}\veps-\frac{k+3}{2} \delta\right)}{2^{k+8-(k+3)\veps-(k+2)\delta} \pi^{\frac{7}{2}-\frac{k+4}{2}\veps-\frac{k+3}{2}\delta}\lsp\Gamma\!\left(1+\frac{\delta}{2}\right) \Gamma\!\left(\frac{3}{2}-\frac{k+3}{2}\veps-\frac{k+2}{2} \delta \right)},
\end{equation}
and
\begin{equation}\label{dkcoeff}
 d_k= \frac{(-1)^k\Gamma\!\left(-\frac{\veps+\delta}{2}\right) \Gamma\!\left(\frac{\veps+\delta}{2}\right)^{k+1} \Gamma\!\left(1-\frac{k+1}{2}\veps-\frac{k}{2} \delta\right) \Gamma\!\left(1-\frac{k+2}{2}\veps-\frac{k+1}{2} \delta\right) \Gamma\!\left(1-\frac{k+3}{2}\veps-\frac{k+2}{2} \delta\right)}{ 2^{k+8-(k+2) \veps-(k+1)\delta} \pi^{\frac{7}{2}-\frac{k+4}{2}\veps-\frac{k+3}{2}\delta}\lsp\Gamma\!\left(1+\frac{\delta}{2}\right) \Gamma\!\left(\frac{3}{2}-\frac{k+2}{2}\veps-\frac{k+1}{2} \delta\right)}\,.
\end{equation}

These expressions are then used to compute the remaining integrals in \eqref{diagrams1sc} utilising the very same techniques as for the integrals in \eqref{diagram4} above. For each bubble, the diagrams in \eqref{diagrams1sc} have one infinite sum that is difficult to perform in closed form. Still, the integrals involved in \eqref{diagrams1sc} are computed term by term. This reduces the renormalisation process to a completely algorithmic procedure that can be implemented in \emph{Mathematica}.

We split the contributions to diagrams (b), (c), and (d) in \eqref{diagrams1sc}, into two parts: one where the bulk interaction point is integrated close to the defect, and one where it is integrated over a distance greater than some finite value. The latter is already made finite by \eqref{Zrenfreebulk} and by the wavefunction renormalisation of $\phi^2$ in the bulk theory. For this reason, this contribution can be neglected, since at this order we only need to keep $\text{O}(\lambda^0 h^k)$ finite terms and $\text{O}(\lambda \lsp h^k)$ divergent terms. In particular, this justifies the use of the expansion \eqref{cd} in diagram (c). Finally, it is easy to reintroduce factors associated with sums over indices: they are just $1$ for diagram (a) and $(N+2)/3$ for diagrams (b), (c), and (d).

Keeping explicit all the sums of each bubble, the integrals in \eqref{diagrams1sc} (with $|x_\perp|=1$), accounting also for symmetry factors are as follows. Diagram (\ref{diagrams1sc}a) is already given in \eqref{onepointallorderapp} and \eqref{bkcoeff}.

For diagram (\ref{diagrams1sc}b)
\begin{equation}\label{diagb}
\begin{tikzpicture}[baseline=7mm,scale=0.5]
 \draw[very thick,black] (-2,0)--(2,0);
 \draw[thick, dashed] (0,3) to[out=-20,in=20](0,1.5);
 \draw[thick, dashed] (0,3) to[out=-160,in=160](0,1.5);
 \draw[thick, dashed] (0,1.5) to[out=-20,in=20](0,0);
 \draw[thick, dashed] (0,1.5) to[out=-160,in=160](0,0);
 \filldraw[color=black,fill=white,line width=0.35mm] (0,1.5) circle (4pt);
 \draw[fill= white] (0,0) circle (15pt);
 \draw[fill= white, draw=black, line width=0.25mm, pattern=north east lines] (0,0) circle (15pt);
 \node[above] at (0,3) {$\phi^2$};
\end{tikzpicture}
\begin{aligned}
  = 3\lambda_0\sum_{k=0}^\infty & \ \frac{(-h_0)^{k+1} \Gamma\! \left(1-\frac{\veps}{2}\right) \Gamma\! \left(1-\veps -\frac{\delta }{2}\right) \Gamma\! \left(\frac{\veps+\delta}{2}\right)^k \Gamma\!\left(-\frac{k+1}{2}\veps-\frac{k}{2}\delta\right)}{2^{k+6-(k+2)\veps-k\delta} \pi^{1-\frac{k+3}{2} \veps-\frac{k+1}{2}\delta}\Gamma \!\left(1+\frac{\delta }{2}\right) \Gamma\! \left(\frac{3}{2}-\frac{\veps}{2}\right)}\\
  &\times \frac{\Gamma\! \left(1-\frac{k}{2}\veps-\frac{k-1}{2}\delta\right)\Gamma\! \left(1-\frac{k+2}{2}\veps-\frac{k+1}{2}\delta\right) }{\Gamma\! \left(1-\frac{\veps+\delta}{2}\right) \Gamma\!\left(\frac{3}{2}-\frac{k+1}{2}\veps-\frac{k}{2}\delta\right)}\,.
\end{aligned}
\end{equation}
For diagram (\ref{diagrams1sc}c)
\begin{equation}\label{diagc}
\begin{aligned}
&\begin{tikzpicture}[scale=0.5, baseline=7mm]
	 \draw[very thick,black] (-3,0)--(3,0);
	 \draw[thick, dashed] (0,3)--(1.5,0);
	 \draw[thick, dashed] (0,3)--(-1,1.5);
  \draw[thick, dashed] (1.5,0)to[out=150,in=-30](-1,1.5);
  \draw[thick, dashed] (-1.5,0)to[out=120,in=210](-1,1.5);
  \draw[thick, dashed] (-1.5,0)to[out=45,in=285](-1,1.5);
  \filldraw[color=black,fill=white,line width=0.35mm] (-1,1.5) circle (4pt);
  \draw[fill= white] (-1.5,0) circle (15pt);
  \draw[fill= white, draw=black, line width=0.25mm, pattern=north east lines] (-1.5,0) circle (15pt);
  \draw[fill= white] (1.5,0) circle (15pt);
  \draw[fill= white, draw=black, line width=0.25mm, pattern=north east lines] (1.5,0) circle (15pt);
	 \node[above] at (0,3) {$\phi^2$};
\end{tikzpicture}
\begin{aligned}
=3\lambda_0\sum_{k_1=0}^\infty \sum_{k_2=0}^\infty&\ \frac{(-h_0)^{k_1+k_2+2} \Gamma\! \left(\frac{\veps+\delta}{2}\right)^{k_1+k_2} \Gamma\! \left(1-\frac{k_1}{2}\veps-\frac{k_1-1}{2}\delta\right)}{2^{6+k_1+k_2- (k_1+k_2+2)\veps- (k_1+k_2)\delta} \pi^{1-\frac{k_1+k_2+4}{2}\veps-\frac{k_1+k_2+2}{2}\delta}}\quad\\
  &\times\frac{\Gamma\! \left(1-\frac{k_2+2}{2}\veps-\frac{k_2+1}{2}\delta\right) \Gamma\! \left(1-\frac{k_2+1}{2}\veps-\frac{k_2}{2}\delta \right)}{\Gamma\! \left(1+\frac{\delta }{2}\right)^2\Gamma\! \left(1-\frac{\veps+\delta }{2}\right) \Gamma\! \left(\frac{3}{2}-\frac{k_1+1}{2}\veps-\frac{k_1}{2}\delta\right)}
\end{aligned}\\
  &\times\bigg(\frac{2^{\veps+\delta}\Gamma\! \left(\tfrac{\veps+\delta }{2}\right) \Gamma\! \left(1-\tfrac{k_1+2}{2}\veps-\tfrac{k_1+1}{2}\delta\right) \Gamma\! \left(-\tfrac{k_1+1}{2}\veps-\tfrac{k_1}{2}\delta\right) \Gamma\!\left(1-\tfrac{k_2+3}{2}\veps-\tfrac{k_2+2}{2}\delta\right)}{\Gamma\! \left(\frac{3}{2}-\frac{k_2+2}{2}\veps-\frac{k_2+1}{2}\delta\right)} \\
  &\quad\quad+\frac{\Gamma\! \left(-\tfrac{\veps+\delta }{2}\right) \Gamma\! \left(-\tfrac{k_1+2}{2}\veps-\tfrac{k_1+1}{2}  \delta\right) \Gamma\! \left(1-\tfrac{k_1+1}{2}\veps-\tfrac{k_1}{2}\delta\right)\Gamma\! \left(1-\tfrac{k_2}{2}\veps-\tfrac{k_2-1}{2}\delta\right)}{\Gamma\!\left(\frac{3}{2}-\frac{k_2+1}{2}\veps-\frac{k_2}{2}\delta\right)}\bigg)\,.
\end{aligned}
\end{equation}
For diagram (\ref{diagrams1sc}d), after a suitable shift in the summation indices one finds
\begin{equation}\label{diagd}
\begin{aligned}
&  \begin{tikzpicture}[scale=0.5, baseline=7mm]
	 \draw[very thick,black] (-3,0)--(3,0);
	 \draw[thick, dashed] (0,3)--(1.5,0);
	 \draw[thick, dashed] (0,3)--(-1.5,0);
  \draw[thick, dashed] (-1.5,0)to(0,1.5);
  \draw[thick, dashed] (1.5,0)to(0,1.5);
  \draw[thick, dashed] (0,1.5)to[out=-60,in=45](0,0);
  \draw[thick, dashed] (0,1.5)to[out=240,in=135](0,0);
  \draw[fill= white] (1.5,0) circle (15pt);
  \draw[fill= white, draw=black, line width=0.25mm, pattern=north east lines] (1.5,0) circle (15pt);
  \draw[fill= white] (0,0) circle (15pt);
  \draw[fill= white, draw=black, line width=0.25mm, pattern=north east lines] (0,0) circle (15pt);
  \filldraw[color=black,fill=white,line width=0.35mm] (0,1.45) circle (4pt);
  \draw[fill= white] (-1.5,0) circle (15pt);
  \draw[fill= white, draw=black, line width=0.25mm, pattern=north east lines] (-1.5,0) circle (15pt);
	 \node[above] at (0,3) {$\phi^2$};
\end{tikzpicture}
\begin{aligned}
  &= -3\lambda_0\sum_{k_1=0}^\infty \sum_{k_2=0}^\infty \sum_{k_3=k_1+k_2}^\infty\, \frac{(-h_0)^{k_3+3} \Gamma\!\left(\frac{\veps+\delta}{2}\right)^{k_3}}{2^{7+k_3-(k_3+3)\veps-(k_3+1)\delta} \pi^{-\frac{3}{2}-\frac{k_3+5}{2}\veps-\frac{k_3+3}{2}\delta}}\\
  &\qquad\times\frac{\Gamma\! \left(1-\frac{k_2}{2}\veps-\frac{k_2-1}{2}\delta\right) \Gamma\!\left(\frac{k_2+3}{2}\veps+\frac{k_2+2}{2}\delta\right) }{\Gamma\!\left(1-\frac{\veps+\delta}{2}\right)\sin\! \left(\pi\left(\frac{k_2+2}{2}\veps+\frac{k_2+1}{2}\delta\right)\right)\sin\!\left(\pi\left(\frac{k_2+1}{2}\veps+\frac{k_2}{2}\delta\right)\right)}
\end{aligned}\\
&\qquad\times\frac{\Gamma\!\left(1-\frac{k_3+3}{2}\veps-\frac{k_3+1}{2}\delta\right) \Gamma\! \left(1-\frac{k_3+5}{2}\veps-\frac{k_3+3}{2}\delta\right) \Gamma\!\left(1-\frac{k_3+4}{2}\veps-\frac{k_3+2}{2} \delta\right)}{\Gamma\!\left(1+\frac{\delta}{2}\right)^2 \Gamma\!\left(\frac{1}{2}+\frac{k_2+2}{2}\veps+\frac{k_2+1}{2}\delta\right) \Gamma\! \left(\frac{3}{2}-\frac{k_2+1}{2}\veps-\frac{k_2 }{2}\delta\right)\Gamma\! \left(\frac{3}{2}-\frac{k_3+4}{2}\veps-\frac{k_3+2}{2} \delta\right)}\,.
\end{aligned}
\end{equation}
Note that $k_1$ does not appear explicitly in the terms of the sum, so its sum can be replaced with a combinatorial factor to speed up calculations: $\sum_{k_1=0}^\infty\sum_{k_2=0}^\infty\sum_{k_3=k_1+k_2}^\infty \to \sum_{k_2=0}^\infty\sum_{k_3=k_2}^\infty (k_3-k_2+1)$.

From these expressions, one can set $h_0 = \mu^{\veps+\delta} Z_{h,\lambda} \, h$ and extract the counterterms $Z_{h,\lambda}$, and then compute the beta function by imposing $\mu\frac{d}{d\mu}h_0=0$. It is
\begin{equation}\label{fullbetah}
\begin{split}
    \beta_h=-( \varepsilon+\delta)h +h^2+\lambda \, (N+2)  \sum_{k=1}^\infty  \beta_{1,k} h^k+\text{O}(\lambda^2,\lambda\veps)\,.
\end{split}
\end{equation}
Since at the bulk fixed point $\lambda_* \propto \veps$, at this order we can just keep the order $\veps^0$ term in the coefficients $ \beta_{1,k}$.\footnote{In the dimensional regularisation with two regulators that we are using, the beta function coefficients are ratios of homogenous polynomials of the same degree in the two regulators (see e.g.\ \eqref{Zphiapp}). At order $\veps^0$, this reduces to the numbers below.
}
The coefficients up to order $k=11$ are
\begin{align}
\beta_{1,1}&= 1\,, \qquad \beta_{1,2}= -2\,, \qquad \beta_{1,3}= 2\,,
\qquad
\beta_{1,4}=-3 + \tfrac{1}{2}\lsp\zeta(3)\,,
\nonumber\\
\beta_{1,5}&= \tfrac{14}{3} - \tfrac{2}{3}\lsp\zeta(3) - \tfrac{3}{4}\lsp\zeta(4) \,,
\qquad
\beta_{1,6}= -\tfrac{15}{2} +\lsp\zeta(3) + \tfrac{9}{8}\lsp\zeta(4) + \tfrac{9}{8}\lsp\zeta(5) \,, \nonumber\\
\beta_{1,7}&=\tfrac{62}{5}-\tfrac85\lsp\zeta(3) -\tfrac95 \lsp\zeta(4)-\tfrac95\lsp\zeta(5) + \tfrac{1}{10}\lsp\zeta(3)^2 - \tfrac{27}{16} \lsp\zeta(6)\,,\nonumber\\
\beta_{1,8}&= -21 +\tfrac83\lsp\zeta(3) +3 \lsp\zeta(4) + 3 \lsp\zeta(5)- \tfrac16 \lsp\zeta(3)^2+ \tfrac{45}{16} \lsp\zeta(6) - \tfrac{3}{8} \lsp\zeta(3)\lsp\zeta(4)+ \tfrac{83}{32} \lsp\zeta(7) \,,\nonumber\\
\beta_{1,9}&= \tfrac{254}{7} -\tfrac{32}{7}\lsp\zeta(3)- \tfrac{36}{7} \lsp\zeta(4) - \tfrac{36}{7} \lsp\zeta(5) +\tfrac27\lsp\zeta(3)^2- \tfrac{135}{28} \lsp\zeta(6) +\tfrac{9}{14}\lsp\zeta(3)\lsp\zeta(4)- \tfrac{249}{56} \lsp\zeta(7)  \nonumber\\
&\hspace{11cm} +\tfrac{9}{14}\lsp\zeta(3)\lsp\zeta(5)- \tfrac{117}{32} \lsp\zeta(8)\,,\nonumber\\
\beta_{1,10}&=-\tfrac{255}{4}+8\lsp\zeta(3)+9\lsp\zeta(4)+ 9 \lsp\zeta(5)- \tfrac12 \lsp\zeta(3)^2+ \tfrac{135}{16} \lsp\zeta(6) -\tfrac98\lsp\zeta(3)\lsp\zeta(4)+ \tfrac{249}{32} \lsp\zeta(7) \nonumber\\
&\hspace{5cm}  -\tfrac98\lsp\zeta(3)\lsp\zeta(5)+ \tfrac{819}{128} \lsp\zeta(8) + \tfrac{1}{48} \lsp\zeta(3)^3- \tfrac{81}{64} \lsp\zeta(4)\lsp\zeta(5)+ \tfrac{2515}{384}\lsp\zeta(9)\,,\nonumber\\
\beta_{1,11}&= \tfrac{1022}{9}-\tfrac{128}{9}\lsp\zeta(3)- 16 \lsp\zeta(4) - 16 \lsp\zeta(5)+\tfrac{8}{9}\lsp\zeta(3)^2- 15 \lsp\zeta(6)+2\lsp\zeta(3)\lsp\zeta(4)- \tfrac{83}{6} \lsp\zeta(7)\nonumber\\
&\hspace{2cm} +2\lsp\zeta(3)\lsp\zeta(5)- \tfrac{91}{8} \lsp\zeta(8)-\tfrac{1}{27} \lsp\zeta(3)^3 + \tfrac94 \lsp\zeta(4)\lsp\zeta(5)+ \tfrac{15}{8}\lsp\zeta(3)\lsp\zeta(6) - \tfrac{2515}{216} \lsp\zeta(9) \nonumber\\
&\hspace{5.5cm} -\tfrac18 \lsp\zeta(3)^2\lsp\zeta(4)+\tfrac98 \lsp\zeta(5)^2+ \tfrac{83}{48}\lsp\zeta(3)\lsp\zeta(7)- \tfrac{2149}{256} \lsp\zeta(10)\,.
\end{align}
The beta function is used to compute the fixed point coupling
\begin{equation}\label{shorth*}
h_*=\delta+\veps
-\frac{N+2}{N+8}\lsp\veps\sum_{k=0}^\infty\beta_{1,k+1}\delta^k\,.
\end{equation}

We turn now to the diagrams contributing to $\langle \hphi (0) \hphi(\tau)\rangle$. At first order in $\lambda$ and to all orders in $h$, they are
\begin{equation}\label{diagram-dtd}
\begin{tikzpicture}[scale=0.5, baseline=0 cm]
 \draw[very thick,black] (-3.5,0)--(3.5,0);
 \draw[thick, dashed] (-3,0)to[out=90,in=180](-1.5,1.5)to[out=0,in=90](0,0);
 \draw[thick, dashed] (3,0)to[out=90,in=0](1.5,1.5)to[out=180,in=90](0,0);
 \draw[fill= white] (0,0) circle (15pt);
 \draw[fill= white, draw=black, line width=0.25mm, pattern=north east lines] (0,0) circle (15pt);
 \node[below] at (0,-1) {(e)};
 \node[below] at (-3,0) {$\hphi$};
 \node[below] at (3,0) {$\hphi$};
\end{tikzpicture}
\qquad
\begin{tikzpicture}[scale=0.5, baseline=0 cm]
 \draw[very thick,black] (-4,0)--(4,0);
 \draw[thick, dashed] (-1.5,0)to[out=90,in=180](0,1.5);
 \draw[thick, dashed] (1.5,0)to[out=90,in=0](0,1.5);
 \draw[thick, dashed] (-3.5,0)to[out=90,in=180](-2.5,1)to[out=0,in=90](-1.5,0);
 \draw[thick, dashed] (3.5,0)to[out=90,in=0](2.5,1)to[out=180,in=90](1.5,0);
 \node[below] at (0,-1) {(f)};
 \draw[thick, dashed] (0,1.5)to[out=-20,in=20](0,0);
 \draw[thick, dashed] (0,1.5)to[out=-160,in=160](0,0);
 \draw[fill= white] (0,0) circle (15pt);
 \draw[fill= white, draw=black, line width=0.25mm, pattern=north east lines] (0,0) circle (15pt);
 \draw[fill= white] (-1.5,0) circle (15pt);
 \draw[fill= white, draw=black, line width=0.25mm, pattern=north east lines] (-1.5,0) circle (15pt);
 \draw[fill= white] (1.5,0) circle (15pt);
 \draw[fill= white, draw=black, line width=0.25mm, pattern=north east lines] (1.5,0) circle (15pt);
 \node[below] at (-3.5,0) {$\hphi$};
 \node[below] at (3.5,0) {$\hphi$};
 \filldraw[color=black,fill=white,line width=0.35mm] (0,1.5) circle (4pt);
\end{tikzpicture}
\end{equation}
Diagram (e) is already computed in~\eqref{effpropapp}.
For (f), we rely on the immediate generalisation of $I'$ as defined in \eqref{diagramApp} and find (at $\tau=1$)
\begin{align}\label{diagf}
&\begin{tikzpicture}[scale=0.5, baseline=0 cm]
  \draw[very thick,black] (-4,0)--(4,0);
  \draw[thick, dashed] (-1.5,0)to[out=90,in=180](0,1.5);
  \draw[thick, dashed] (1.5,0)to[out=90,in=0](0,1.5);
  \draw[thick, dashed] (-3.5,0)to[out=90,in=180](-2.5,1)to[out=0,in=90](-1.5,0);
  \draw[thick, dashed] (3.5,0)to[out=90,in=0](2.5,1)to[out=180,in=90](1.5,0);
  \draw[thick, dashed] (0,1.5)to[out=-20,in=20](0,0);
  \draw[thick, dashed] (0,1.5)to[out=-160,in=160](0,0);
  \draw[fill= white] (0,0) circle (15pt);
  \draw[fill= white, draw=black, line width=0.25mm, pattern=north east lines] (0,0) circle (15pt);
  \draw[fill= white] (-1.5,0) circle (15pt);
  \draw[fill= white, draw=black, line width=0.25mm, pattern=north east lines] (-1.5,0) circle (15pt);
  \draw[fill= white] (1.5,0) circle (15pt);
  \draw[fill= white, draw=black, line width=0.25mm, pattern=north east lines] (1.5,0) circle (15pt);
  \node[below] at (-3.5,0) {$\hphi(0)$};
  \node[below] at (3.5,0) {$\hphi(1)$};
  \filldraw[color=black,fill=white,line width=0.35mm] (0,1.5) circle (4pt);
\end{tikzpicture}
= -3\lambda_0\sum_{k_1=0}^\infty\sum_{k_2=0}^\infty\sum_{k_3=k_1+k_2}^\infty \frac{(-h_0)^{k_3+1} \Gamma\! \left(\frac{\veps+\delta }{2}\right)^{k_3}}{ 2^{4+k_3+\veps+\delta} \pi^{-1-\frac{k_3+3}{2}\veps-\frac{k_3+1}{2}\delta}} \nonumber\\
&\qquad\qquad\times \frac{\Gamma\! \left(\frac{k_2+3}{2}\veps+\frac{k_2+2}{2}\delta\right) \Gamma\! \left(1-\frac{k_2}{2}\veps-\frac{k_2-1}{2}\delta\right)}{\Gamma\!\left(1+\frac{\delta }{2}\right) \Gamma\!\left(1-\frac{\veps+\delta}{2}\right) \sin\!\left(\pi\left(\frac{k_2+2}{2}\veps+\frac{k_2+1}{2}\delta\right)\right) \sin\!\left(\pi\left(\frac{k_2+1}{2}\veps+\frac{k_2}{2}\delta\right)\right)} \\
&\qquad\qquad\times \frac{\Gamma\! \left(1-\frac{k_3+3}{2}\veps-\frac{k_3+1}{2}\delta\right)}{\Gamma\! \left(\frac{1}{2}+\frac{k_2+2}{2} \veps+\frac{k_2+1}{2}\delta\right) \Gamma\! \left(\frac{3}{2}-\frac{k_2+1}{2}\veps-\frac{k_2}{2}\delta\right) \Gamma\! \left(\frac{k_3+3}{2}\veps+\frac{k_3+2}{2}\delta\right)}\,.
\nonumber
\end{align}
As in \eqref{diagd}, $k_1$ does not appear in the sum, so one can reduce the expression to a double sum.

From these diagrams we extract the counterterms contributing to the wavefunction renormalisation of $\hphi^a_0= Z_{\hphi} \,\hphi^a$; they are
\begin{align}\label{Zphiapp}
Z_{\hphi} &= 1-\frac{h }{\veps+\delta} +\lambda\, \frac{N+2}{2\veps+\delta} \bigg( h-\frac{3 \veps +\delta}{3 \veps +2 \delta}h^2\nonumber\\
&\quad+\frac{8(\veps+\delta)(5\veps+2\delta)-\veps\delta\zeta(2)-4(\veps+\delta)(2\veps+\delta)\zeta(3)}{4(3\veps+2 \delta) (4\veps+3\delta)}h^3 \nonumber\\
&\quad -\frac{ 1}{4(3 \veps + 2 \delta ) (4 \veps + 3 \delta) (5 \veps+ 4 \delta )} \Big(8 (\veps +\delta)^2(25 \veps +11\delta)+ \veps \delta  (\veps-\delta)\zeta (2)\\
&\qquad\qquad  -\left(32 \veps^3+81  \veps ^2\delta +66  \veps \delta ^2 +16 \delta^3\right)\zeta (3) -18(\veps +\delta )^2(2 \veps+\delta)\zeta (4)\Big) h^4
\nonumber\\
&\quad + \text{O}(h^5) \bigg) + \text{O}(\lambda^2) \, + \, \text{higher order poles} \,.
\nonumber
\end{align}
We omit higher order poles since they can be easily reconstructed using the 't~Hooft relations. It is not necessary to compute more terms, since it is already easy to guess the series for the anomalous dimension of $\hphi$ at the fixed point,
\begin{equation}
\label{gammaphi}
   \gamma_{\hphi}=\left[\frac{d \log Z_{\hphi}}{d h}\beta_h+\frac{d \log Z_{\hphi}}{d \lambda}\beta_\lambda\right]_{\lambda_*,h_*} \hspace{-0.3cm}=\delta+\veps-\frac{N+2}{N+8} \lsp\veps\sum_{k=0}^\infty (-\delta)^k  + \text{O}(\veps^2)\,.
\end{equation}
Using this anomalous dimension one finds \eqref{dimphiint}.

Finally, the dimension $\hat{\Delta}_{\del_\perp \hphi}$ of the defect operator $\del_\perp \hphi$ is computed from the defect two-point function $\langle \del_\perp \hphi(0) \del_\perp \hphi(\tau)\rangle $. Up to order one in $\lambda$ and at all orders in $h$ is given by the diagrams\footnote{Note that a single normal derivative of the free defect-to-defect propagator $\del_\perp G(x-y) \big|_{x_\perp=y_\perp=0}$ vanishes, so extra bubbles as in \eqref{diagram-dtd} do not contribute.\label{foot:deriv}} (with $SO(d-p)$ indices of $\del_\perp \hphi$ indices contracted)
\begin{align}\label{diagg}
\begin{tikzpicture}[baseline=(vert_cent.base),scale=0.5]
  \node[inner sep=0pt,outer sep=0pt] (vert_cent) at (0,0) {$\phantom{\cdot}$};
  \draw[very thick,black] (-3,0)--(3,0);
  \draw[thick, dashed, shorten <=2] (-2,0)to[out=90,in=180](0,1.5)to[out=0,in=90](2,0);
  \node[below] at (-2,0) {$\del_\perp \hphi(0)$};
  \node[below] at (2,0) {$\del_\perp \hphi(1)$};
\end{tikzpicture}\!\!
&= \frac{(2-\veps-\delta) \Gamma\! \left(2-\frac{\veps}{2}\right)}{2 \pi ^{2-\frac{\veps}{2}}} \,,
\\\label{diagh}
\begin{tikzpicture}[scale=0.5, baseline=(vert_cent.base)]
  \node[inner sep=0pt,outer sep=0pt] (vert_cent) at (0,0) {$\phantom{\cdot}$};
  \draw[very thick,black] (-3,0)--(3,0);
  \draw[thick, dashed] (-2,0)to[out=90,in=180](0,1.5);
  \draw[thick, dashed] (2,0)to[out=90,in=0](0,1.5);
  \draw[thick, dashed] (0,1.5)to[out=-20,in=20](0,0);
  \draw[thick, dashed] (0,1.5)to[out=-160,in=160](0,0);
  \draw[fill= white] (0,0) circle (15pt);
  \draw[fill= white, draw=black, line width=0.25mm, pattern=north east lines] (0,0) circle (15pt);
  \node[below] at (-2,0) {$\del_\perp \hphi(0)$};
  \node[below] at (2,0) {$\del_\perp \hphi(1)$};
  \filldraw[color=black,fill=white,line width=0.35mm] (0,1.5) circle (4pt);
\end{tikzpicture}\!\!
&=-3\lambda_0\sum_{k=0}^\infty \frac{(-h_0)^{k+1} ((k+1)\veps+k\delta) \Gamma\! \left(\frac{\veps+\delta }{2}\right)^k }{ 2^{3+k+\veps+\delta } \pi ^{-1-\frac{k+3}{2}\veps-\frac{k+1}{2} \delta} \Gamma\! \left(1+\frac{\delta }{2}\right) \Gamma\! \left(1-\frac{\veps+\delta }{2}\right)}
\\\nonumber
&\hskip-2.5cm\times \frac{\Gamma\! \left(1-\frac{k}{2}\veps-\frac{k-1}{2}\delta\right) \Gamma\! \left(2-\frac{k+3}{2}\veps-\frac{k+1}{2}\delta\right)}{\sin\! \left(\pi\left(\frac{k+2}{2}\veps+\frac{k+1}{2} \delta\right)\right)\sin\! \left(\pi\left(\frac{k+1}{2}\veps+\frac{k}{2}\delta\right)\right)\Gamma\! \left(\frac{1}{2}+\frac{k+2}{2}\veps+\frac{k+1}{2}\delta\right) \Gamma\! \left(\frac{3}{2}-\frac{k+1}{2}\veps-\frac{k}{2}\delta\right)}\,.
\end{align}
The simple poles of the wavefunction renormalisation of $\del_\perp \hphi^a_0=Z_{\del_\perp \hphi}\, \del_\perp \hphi^a$ are
\begin{align}\label{Zdelphi}
Z_{\del_\perp \hphi}&= 1 + \lambda\,\frac{N+2}{2\veps + \delta} \Big( \frac{h}{2}-\frac{ \veps +\delta }{4  (3 \veps +2 \delta )}h^2
+ \frac{ 6 (\veps +\delta )^2+\veps\delta \zeta(2)}{8  (3 \veps +2 \delta ) (4 \veps +3 \delta )}h^3
\nonumber\\
     &\quad-\frac{30 (\veps +\delta )^3+3\veps\delta(\veps +\delta )\zeta(2)+2 \veps \delta (2\veps+\delta )\zeta(3)}{16 (3 \veps +2 \delta ) (4 \veps +3 \delta ) (5 \veps +4 \delta )}h^4\nonumber\\
     &\quad+ \frac{528 (\veps +\delta )^4+72    \veps \delta (\veps +\delta )^2\zeta (2)+16   \veps\delta (\veps+\delta)(\veps+2 \delta)\zeta (3)}{64(3 \veps +2 \delta ) (4 \veps +3 \delta ) (5 \veps +4 \delta ) (6 \veps +5 \delta )}h^5\nonumber\\
     &\quad+ \frac{3\veps \delta  \left(4 \veps ^2+17 \delta  \veps+ 12 \delta ^2\right)\zeta (4)}{64  (3 \veps +2 \delta ) (4 \veps +3 \delta ) (5 \veps +4 \delta ) (6 \veps +5 \delta )}h^5
     + \text{O}(h^6)\Big) + \text{O}(\lambda^2)\, \nonumber\\
      &\quad+ \ \text{higher order poles} \,.
\end{align}
From these terms it is already possible to guess the series for the anomalous dimension of $\del_\perp \hphi$ at the fixed point
\begin{equation}\label{delphiresum}
   \gamma_{\del_\perp \hphi}=\left[\frac{d \log Z_{\del_\perp\hphi}}{d h}\beta_h+\frac{d \log Z_{\del_\perp \hphi}}{d \lambda}\beta_\lambda\right]_{\lambda_*,h_*} \hspace{-0.3cm}= -\frac{N+2}{N+8} \lsp\veps\sum_{k=1}^\infty \frac{\delta^k-(-2\delta)^k}{3 \cdot 2^k} + \text{O}(\veps^2)\,.
\end{equation}
Using this anomalous dimension one finds \eqref{dimdelphi}.

\section{Higher order results for surface defects}
\label{sec:deg}
In this appendix we set $\delta=0$ and discuss surface defects in $d=4-\veps$. These have been studied already in \cite{Trepanier:2023tvb, Raviv-Moshe:2023yvq, Giombi:2023dqs}, and it was found that there also exist symmetry-breaking conformal surface defects which realise the subgroup $O(m)\times O(n)$ of the bulk $O(N)$ symmetry, with $m+n=N$. The discussion in those works was limited to leading order results, and here we extend it to next-to-leading order.

To evaluate the beta function of surface defect couplings $h_{ab}$ at next-to-leading order, terms of order $\lambda h^2$ can be found using results in the main text, specifically \eqref{interactingbeta}. Terms of order $\lambda^2 h$ can be computed using a trick discussed in~\cite{Harribey:2024gjn}; since they are linear in $h$, they are easily obtained from the dimensions of corresponding operators in the bulk theory (at the trivial defect). The result is
\begin{equation}\label{eq:betaSurfaceNLO}
    \beta_{ab}=-\varepsilon\lsp h_{ab}+h_{ac}h_{bc}+\lambda[2\lsp h_{ab}+h_{cc}\delta_{ab}-2(2\lsp h_{ac}h_{bc}+h_{ab}h_{cc})]-\tfrac16\lambda^2[(N+10)h_{ab}+4\lsp h_{cc}\delta_{ab}]\,.
\end{equation}

For the symmetry preserving case, $\beta_{ab}= \delta_{ab} h$, \eqref{eq:betaSurfaceNLO} becomes
\begin{equation}
    \beta_h= - \veps h + h^2 + (N+2) \lambda h \big(1-2h-\tfrac56\lambda\big)\,,
\end{equation}
which agrees with the result reported in \cite{Diatlyk:2024ngd}. With $\lambda_*=\frac{\veps}{N+8}+\frac{3(3N+14)}{(N+8)^3}\veps^2$ \cite{Kleinert:2001ax}, the dimension of $\hat{S}$ at the non-trivial fixed point is
\begin{equation}\label{surfacenexttoleadingdimension}
    \hD_{\hat{S}}=2+\frac{\del \beta_h}{\del h}\Big|_{\lambda_*,h_*}\!\!=2+ \frac{6 \lsp\veps}{N+8}-\frac{(N+2)(13N+44)\,\veps^2 }{2(N+8)^3}+\text{O}(\veps^3)\,.
\end{equation}
In the similar context of the $\phi^2$ deformation integrated on a boundary, the extrapolation from the two-loop result \cite{Diehl:1998mh} was already very close to the result of the numerical boundary bootstrap \cite{Gliozzi:2015qsa}. We can expect a similar behaviour here. Indeed, for example by setting $N=1$ we find that the Padé approximants at $\veps=1$ are already relatively close to each other
\begin{equation}\label{dimensioninterfaceoperator}
    \hD_{\hat{S}} \lsp[2/0] \approx 2.55 \,, \qquad \hD_{\hat{S}} \lsp[1/1] \approx  2.57\,.
\end{equation}
However, a singlet primary operator with such dimension should be absent if the fixed point is an interface with Dirichlet boundary conditions, as proposed in \cite{Krishnan:2023cff}. Numerical evidence presented in \cite{Diatlyk:2024ngd} suggests that the operator $\hat{S}$ factorises into two different copies of the boundary operator $\del_\perp  \hphi$. Indeed one can check that $\hat{\Delta}_{\hat{S}}\approx2\lsp\hat{\Delta}_{\del_\perp\hphi}$.

Moving on to the symmetry-breaking case, for $O(m)\times O(n)$ symmetry with $m+n=N$ there are two couplings $h_m,h_n$ with beta functions that follow directly from \eqref{eq:betaSurfaceNLO}
\begin{align}
    \beta_{h_m}&=-\veps\lsp h_m+h_m^2+\lambda[(m+2)h_m+nh_n-2(m+2)h_m^2-2nh_mh_n]\nonumber\\&\hspace{6cm}-\lambda^2[\tfrac12h_m(5m+n+10)+2nh_n]\label{eq:betahm}\,,\\
    \beta_{h_n}&=-\veps\lsp h_n+h_n^2+\lambda[(n+2)h_n+mh_m-2(n+2)h_n^2-2mh_mh_n]\nonumber\\&\hspace{6cm}-\lambda^2[\tfrac12h_n(m+5n+10)+2mh_m]\label{eq:betahn}\,.
\end{align}
Zeroes of these beta functions at the $O(N)$ bulk fixed point, where $\lambda=\frac{1}{N+8}\veps+\frac{3(3N+14)}{(N+8)^3}\veps^2$, are found by substituting $h_m=h_m^{(1)}\veps+h_m^{(2)}\veps^2$ and similarly for $h_n$ into \eqref{eq:betahm}, \eqref{eq:betahn} and solving the resulting equations order by order in $\veps$. At leading order this yields two fixed points, given by
\begin{equation}
    h_{m,\pm}^{(1)}=\frac{m+3\pm\sqrt{9-mn}}{m+n+8}\,,\qquad h_{n,\pm}^{(1)}=\frac{n+3\mp\sqrt{9-mn}}{m+n+8}\,.
\end{equation}
The expressions for $h_{m,\pm}^{(2)}, h_{n,\pm}^{(2)}$ and the analogue of \eqref{surfacenexttoleadingdimension} are easy to work out but too complicated to present here.

Due to the square root $\sqrt{9-mn}$, these fixed points exist only for $mn\leq9$. This inequality is saturated for integer $m,n$ when $m=1,n=9$ and when $m=n=3$ where, without loss of generality, we have taken $m\leq n$. For these values of $m,n$ the two fixed points collide and they move off to the complex plane for $mn>9$. This collision point is corrected at next-to-leading order in $\varepsilon$, and a standard analysis following \cite{Osborn:2017ucf} shows that the critical $n$ for which the two fixed points collide is given by
\begin{equation}
    n_{c}=\frac{9}{m}-\frac{3(m^2-22\lsp m+9)}{2m(m^2+8\lsp m+9)}\lsp\veps+\text{O}(\veps^2)\,.
\end{equation}
For $m=1$ this becomes $9+\veps$ and for $m=3$ it becomes $3+\frac47\veps$. In both cases the collision point lies above $9$ and $3$, respectively, which renders the fixed points at $m=1,n=9$ and $m=n=3$ unitary. Nevertheless, the $\veps$ expansion we have been using so far breaks down and an expansion of $h_m,h_n$ involving half-integer powers of $\veps$ is necessary to describe them. For $m=1, n=9$ we find the two fixed points
\begin{equation}
    h_{m=1,\pm}=\tfrac23\lsp\veps\pm\tfrac{1}{18}\lsp\veps^{3/2}+\text{O}(\veps^2)\,,\qquad h_{n=9,\pm}=\tfrac29\lsp\veps\mp\tfrac{1}{18}\lsp\veps^{3/2}+\text{O}(\veps^2)\,,
\end{equation}
and for $m=n=3$
\begin{equation}
    h_{m=3,\pm}=\tfrac37\lsp\veps\pm\tfrac{\sqrt{21}}{49}\lsp\veps^{3/2}+\text{O}(\veps^2)\,,\qquad h_{n=3,\pm}=\tfrac37\lsp\veps\mp\tfrac{\sqrt{21}}{49}\lsp\veps^{3/2}+\text{O}(\veps^2)\,.
\end{equation}
Let us note that in the $m=n=3$ case the defect symmetry at leading order is $O(3)^2\rtimes \mathbb{Z}_2$, but this is broken to $O(3)^2$ at next-to-leading order.

\end{appendices}

\bibliography{references}

\end{document}